\documentclass[11pt]{article}

\title{
\Large \bf{Hidden gauge reducibility of superstring field theory \\ and Batalin-Vilkovisky master action} 
}\author{\Large{Hiroaki Matsunaga}\footnote{Email: matsunaga@fzu.cz} 
}\date{\it Institute of Physics of the Czech Academy of Sciences, \\ Na Slovance 2, Prague 8, Czech Republic \vspace{-7mm} }

\usepackage[dvips]{graphicx}
\usepackage{amsmath,amscd, amssymb}
\usepackage{cite}
\usepackage{mathrsfs} 
\usepackage{amsfonts}
\usepackage{color} 
\usepackage{upgreek}
\usepackage[hypertex]{hyperref}

\makeatletter
  
  \@addtoreset{equation}{section}
\makeatother

\newcommand{\la}{\big{\langle }}
\newcommand{\ra}{\big{\rangle }} 
\newcommand{\La}{\Big{\langle }}
\newcommand{\Ra}{\Big{\rangle }}
\newcommand{\LA}{\bigg{\langle }}
\newcommand{\RA}{\bigg{\rangle }} 
\newcommand{\lla}{\big{\langle } \hspace{-1.3mm} \big{\langle }} 
\newcommand{\rra}{\big{\rangle } \hspace{-1.3mm} \big{\rangle }} 
\newcommand{\srra}{\rangle \hspace{-1mm} \rangle } 
\newcommand{\slla}{\langle \hspace{-1mm} \langle }

\newcommand{\no}{\nonumber\\ }

\newcommand{\sub}[1]{\vspace{-8pt}\subsubsection*{#1}\vspace{-4pt}}

\newcommand{\Eta}{\boldsymbol{\eta }} 
\newcommand{\bxi }{\boldsymbol{\xi }} 
\newcommand{\blambda}{\boldsymbol{\lambda }} 
\newcommand{\bLambda}{\boldsymbol{\Lambda }} 
 
\newcommand{\bPsi }{\boldsymbol{\Psi }}
\newcommand{\bPhi }{\boldsymbol{\Phi }}
\newcommand{\bM}{\mathbf{M}}

\newcommand{\cB}{\mathcal{B}} 
 
\newcommand{\cD}{\mathcal{D}}

\newcommand{\cG}{\mathcal{G}}
\newcommand{\cH}{\mathcal{H}}
\newcommand{\cI}{\mathcal{I}}

\newcommand{\cL}{\mathcal{L}} 
\newcommand{\cM}{\mathcal{M}}

\newcommand{\cR}{\mathcal{R}}

\newcommand{\cV}{\mathcal{V}} 
\newcommand{\cW}{\mathcal{W}}

\newcommand{\cZ}{\mathcal{Z}} 

\allowdisplaybreaks[3]

\begin{document}
\maketitle
\begin{abstract}
In this paper, we show that there exists a hidden gauge reducibility in superstring field theory based on the small dynamical string field $\Psi \in \cH _{\beta \gamma }$ whose gauge variation is also small $\delta \Psi \in \cH _{\beta \gamma }$. 
It requires additional ghost--antighost fields in the gauge fixed or quantum gauge theory, and thus changes the Batalin-Vilkovisky master action, which implies that additional propagating degrees of freedom appear in the loop superstring amplitudes via the gauge choice of the field theory. 
We present that the resultant master action can take a different and enlarged form, and that there exist canonical transformations getting it back to the canonical form. 
On the basis of the Batalin-Vilkovisky formalism, we obtain several exact results and clarify this underlying gauge structure of superstring field theory. 
\end{abstract}
\vspace{-5mm} 
\setcounter{tocdepth}{2}
{\small \tableofcontents } 
\thispagestyle{empty} 
\addtocounter{page}{-1}

\section{Introduction} 

Gauge theory is a theory whose dynamical variables are redundant, in which we should take gauge degrees of freedom into account and clarify {\it its gauge reducibility} \cite{Batalin:1981jr, Batalin:1984jr, Henneaux:1989jq, Gomis:1994he}. 
Iff all gauge transformations are independent, it is called irreducible, which is the simplest gauge theory. 
In other cases, reducible gauge theory, a kind of the $(g+1)$-st gauge invariance for the $g$-th gauge invariance arises and there exists a hierarchy of gauge invariances. 
If $g$-th gauge transformations are independent, then the theory is called $g$-th order reducible. 

\sub{Infinitely reducible}

It is known that (super-)string field theory is an infinitely reducible gauge theory, and thus it necessitates a set of infinite number of ghost--antighost fields for the quantization. 
These ghost--antighost string fields never appear in the action without gauge fixing, but propagate in the loop amplitudes. 
One can find what kind of and how many ghosts are necessitated by studying the gauge reducibility of the kinetic term, namely, free theory.\footnote{It is equivalent to clarify the existence condition of the propagator in given gauge theory.} 
In many cases, the kinetic operator of (super-)string field theory is the BRST operator $Q$ of the world-sheet theory, which gives the on-shell condition $Q \, \Psi = 0$ of a string field $\Psi$.
Since $Q$ is nilpotent, it has the gauge invariance under $\delta \Psi = Q \, \lambda _{0}$. 
However, clearly, there exists {\it the gauge transformation for the gauge transformation} $\delta _{1} \lambda _{0} = Q \, \lambda _{-1}$, where we write $\lambda _{1-g}$ for the $g$-th gauge parameter field. 
Likewise, we find the $g$-th gauge variation $\delta _{g} \lambda _{1-g} = Q \lambda _{-g}$ preserving $(g-1)$-th gauge invariance. 
\begin{center}
\begin{tabular}{|c||c|c|}
\hline 
e.o.m. & gauge invariance & gauge invariance for gauge invariance 
\\ \hline 
$Q \, \Psi = 0$ & $\delta \, \Psi = Q \, \lambda _{0}$ 
& $\delta _{g} \, \lambda _{1-g} = Q \, \lambda _{-g}$ \hspace{2mm} $(g >0)$  
\\ \hline  
\end{tabular} 
\end{center}
When the set of gauge parameters $\{ \lambda _{-g} \} _{g\geq 0} = \{ \lambda _{0} , \lambda _{-1} , ... , \lambda _{-g}, ... \}$ appears in the analysis of the gauge reducibility, the theory needs corresponding ghost fields $\{ \Psi _{-g} \} _{g\geq 0}$. 
The pair of the string field and ghost fields $\{ \Psi , \Psi _{-g} \} _{g\geq 0}$ requires its antighost fields part $\{ \Psi ^{\ast } , (\Psi _{-g})^{\ast } \} _{g\geq 0}$.    
These fields appear in gauge fixed theory and quantum calculations. 
Hence, in the Batalin-Vilkovisky formalism \cite{Batalin:1981jr, Batalin:1984jr}, {\it the minimal set of fields--antifields} is given by $\{ \Psi , \Psi ^{\ast } , \Psi _{1-g} , (\Psi _{1-g} )^{\ast } \} _{g> 0}$: 
\begin{center}
\begin{tabular}{|c||c|} 
\hline 
Fields & $\Psi $\,, $\{ \Psi _{1-g} \} _{g>0}$
\\ \hline  
\end{tabular} 
\hspace{2mm} 
\begin{tabular}{|c||c|}
\hline 
Antifields & $ (\Psi _{1})^{\ast } \equiv \Psi ^{\ast }$\,, $\{ (\Psi _{1-g})^{\ast } \} _{g>0}$ 
\\ \hline  
\end{tabular} 
\end{center}
One promised way to achieve gauge fixed or quantum gauge theory is to construct a Batalin-Vilkovisky master action $S_{\sf bv}$ based on the minimal set of fields--antifields. 
Usually, it is a tough work to find $S_{\sf bv}$ for given gauge theory: 
In many cases, $S_{\sf bv}$ will need nontrivial ghost--antighost terms in addition to the original action $S$ and be given by a highly complicated form. 
However, in (super-)string field theory, we often encounter an interesting situation: 
The master action $S_{\sf bv}$ takes the same form as the original action $S $ except that it includes all fields--antifields. 
In general, an action $S [\Psi ]$ for interacting (super-)string field theory takes the following form 
\begin{align}
\label{SFT action}
S [ \Psi ] = \overbrace{K( \Psi , \, Q \, \Psi )}^{\rm kinetic \,\, term} 
+ \sum_{n\geq 3} \overbrace{V_{n} ( \Psi , ... , \Psi ) }^{{\rm tree} \,\, n\mathchar`-{\rm vertex}} 
+ \sum_{g} \sum_{n\geq 1} \overbrace{V_{g,n} ( \underbrace{\Psi , ... , \Psi }_{n} ) }^{g\mathchar`-{\rm loop \,\, correction}} \, . 
\end{align}
We know that in several types of (super-)string field theory, such as \cite{Witten:1985cc, Zwiebach:1992ie, Gaberdiel:1997ia, Witten:1986qs, Jurco:2013qra, Erler:2013xta, Erler:2014eba, Sen:2015uaa, Konopka:2016grr, Erler:2016ybs}, one can obtain its classical (and quantum) Batalin-Vilkovisky master action $S_{\sf bv}$ without changing the form of the original action: 
Namely, $S_{\sf bv} = S[\psi ]$ where $\psi \equiv \Psi + \sum_{g<1} \Psi _{g} + \sum_{g\leq 1} (\Psi _{g})^{\ast }$. 
Free theory gives a more trivial example. 
Let us consider the kinetic term $K = K(\Psi , Q \Psi )$ of \cite{Jurco:2013qra, Erler:2013xta, Erler:2014eba, Sen:2015uaa, Konopka:2016grr, Erler:2016ybs, Ohmori:2017wtx},\footnote{There is auxiliary kinetic term $K_{\sf aux}[\Psi ,\widetilde{\Psi }]$ for \cite{Sen:2015uaa, Konopka:2016grr}, and $K_{\sf bv}$ takes the form of $K_{\sf bv} = K[\psi ] + K_{\sf aux} [ \psi , \widetilde{\psi } ]$.} 
\begin{align*}
K \big( \Psi , Q \, \Psi \big) = \frac{1}{2} \lla \Psi , \, Q \, \Psi \rra \, .
\end{align*}
Since this is a free theory, its master action $K_{\sf bv}$ is obtained by adding Lagrange-multiplier-like ghost--antighost terms fixing the above reducibility. 
Thus $K_{\sf bv}$ takes the following form  
\begin{align*}
K_{\sf bv} = \frac{1}{2} \lla \Psi , \, Q \, \Psi \rra + \sum_{g>0} \lla (\Psi _{g})^{\ast } , \, Q \, \Psi _{1-g} \rra 
= K \big( \psi , Q \, \psi \big) \, . 
\end{align*}
Why can we obtain the master action $S_{\sf bv}[\psi ]$ by just relaxing the ghost number constraint of the original action $S[\Psi ]$? 
This naive question is the first motivation of this paper. 
While one may think that it is provided by the nilpotency of $Q$ in the case of free theory (or homotopy algebraical relations satisfied by the set of vertices $\{ V_{g,n} \} _{g,n}$ for interacting theory), it seems that there is an additional reason related to its geometrical interpretation. 
Recall that the (super-)string BRST operator $Q$ works as the exterior derivative on the moduli space $\cM _{g,n}$ of (super-)Riemann surfaces $\Sigma _{g,n}$ with $g$-genus and $n$-punctures.\footnote{Off course, for superstrings, NS and R punctures should be distinguished: $n = (n_{\rm NS} | n_{\rm R})$.} 
For example, see \cite{AlvarezGaume:1988bg, AlvarezGaume:1988sj, Kimura:1993ea, Belopolsky:1997bg}. 

\sub{Geometrical restriction}

It is well-known that the on-shell $g$-loop $n$-point amplitude of (super-)strings are described by the integration over the (super-)moduli space $\cM _{g,n}$. 
String field theory gives its off-shell extension as a gauge theory, and there are several ways to construct gauge invariant actions reproducing this on-shell property. 
However, as we know, a straightforward but powerful way of constructing field theory is to regard it as one of the off-shell defining properties: 
Consider a Feynman graph decomposition\footnote{Here, $\cR ^{(I)}_{g,n}$ denotes the region covered by all $g$-loop $n$-point graphs including $I$-propagators: $\cV _{g,n} \equiv \cR ^{(0)}_{g,n}$.} of the (super-)moduli space $\cM _{g,n} = \cV _{g,n} \cup \cR _{g,n}^{(1)} \cup \dots \cup \cR_{g,n}^{(3g-3+n)}$, find $n$-fold multilinear functions $\omega _{g,n}$ of string fields which are pull back from the volume forms of $\cM _{g,n}$, and define all vertices $V_{g,n}$ of superstring field theory by 
\begin{align*}
V_{g,n} \equiv \int _{\cV _{g,n}} \omega _{g,n} \, . 
\end{align*}
Then, one can prove that $Q$ and the resultant vertices $\{ V_{g,n} \} _{g,n}$ satisfy the (loop) $A_{\infty } / L_{\infty }$ relations, which are key algebraic relations providing a simple Batalin-Vilkovisky procedure.\footnote{By taking advantage of this old well-known fact, we can algebraically construct string tree vertices satisfying $A_{\infty } / L_{\infty }$ relations without references to these geometrical aspects \cite{Erler:2013xta, Erler:2014eba, Erler:2016ybs}. 
Interestingly, this purely algebraic prescription, the homotopy algebraic formulation, also reproduces the same on-shell tree amplitudes \cite{Konopka:2015tta}. 
However, while these vertices satisfy the $A_{\infty } / L_{\infty }$ relations by construction, it has remained unclear whether they give pull-backs of differential forms on $\{ \cM _{g,n} \} _{g,n}$. 
Recently, this point was clarified for lower vertices \cite{Ohmori:2017wtx}.} 
If these $\{ Q , V_{g,n} \} _{g,n}$ give {\it a unique} gauge generator of theory, we can obtain $S_{\sf bv} = S [\psi ]$. 

A string field $\Psi $ should be correspond to a set of world-sheet vertex operators inserted into the coordinate patch around each puncture of $\Sigma _{g,n}$. 
In this set up, it would be simple and natural to use a string field living on the small Hilbert space $\cH _{\beta \gamma }$ because this geometrical interpretation arises from the properties of $bc$- and $\beta \gamma$-ghost systems. 
Roughly, the gauge invariance of this type of string field theory is a result of the Stokes theorem on the set $\{ \cM _{g,n} \} _{g,n}$ of the (super-)moduli spaces. 
Thus, if we want to keep this geometrical interpretation of the gauge invariance off-shell, it would be reasonable to restrict not only the string field $\Psi $ but also its gauge variation $\delta \Psi $ onto the small Hilbert space $\cH _{\beta \gamma }$. 
Using $\Psi \in \cH _{\beta \gamma }$ satisfying $\delta \Psi \in \cH _{\beta \gamma }$ as a dynamical string field, we can obtain an off-shell gauge theory based on these. 

On the basis of the minimal set of fields--antifields $\psi $ which also satisfies $\psi \in \cH _{\beta \gamma }$, one can obtain the master action $S_{\sf bv}$ for this type of gauge theory $S[\Psi ]$ by just relaxing ghost number constraint $S_{\sf bv} = S[\psi ]$, which would be a well-established fact. 

\sub{Unrestriction}

However, there is an implicit but too strong assumption in the derivation of these master actions: All fields--antifields are restricted onto the small Hilbert space $\cH _{\beta \gamma }$. 
There exists a slight mismatch between the statements ``the gauge variation is small $\delta \Psi \in \cH _{\beta \gamma }$'' and ``the gauge parameter field is small $\lambda \in \cH_{\beta \gamma }$'', which becomes significant for higher gauge parameters. 
In the case of free theory, we take the gauge variation $\delta \Psi = Q \, \lambda$. 
Apparently, to be $\delta \Psi \in \cH _{\beta \gamma }$, the gauge parameter $\lambda $ must belong to $\cH _{\beta \gamma }$ except for $Q$-exact terms, and one can find that other higher gauge parameter fields $\lambda _{1-g}$ do not have to be small. 
In other words, although we start from the small dynamical string field $\Psi \in \cH _{\beta \gamma }$ satisfying $\delta \Psi \in \cH _{\beta \gamma }$, the first gauge parameter $\lambda $ can protrude from $\cH _{\beta \gamma }$ as long as BRST exact, $\lambda \in \cH _{\beta \gamma } \oplus Q \cH _{\xi \eta \phi }$, and higher gauge parameters $\lambda _{-g}$ are no longer restricted, $\lambda _{-g} \in \cH _{\xi \eta \phi }$. 
We call this unrestricted state space $\cH _{\xi \eta \phi }$ as the large Hilbert space \cite{Friedan:1985ge}. 
Hence, using the above minimal set of fields--antifields $\psi $ satisfying $\psi \in \cH _{\beta \gamma }$ is too restrictive, which is indeed sufficient but not necessary. 
As a gauge theory, it corresponds to fix some higher gauge symmetry, which we see in the next section. 

What happens if we {\it unrestrict} these constraints on these gauge parameters? 
Clearly, the action $S[\Psi ]$ and its (first) gauge invariance do not change. 
However, we can consider more enlarged gauge transformations for the gauge transformations, the \textit{large} gauge symmetry of the \textit{small} theory, and find quite different gauge hierarchy: 
The gauge reducibility of superstring field theory is drastically changed. 
As we will see, it necessitates {\it additional} ghost and antighost fields in the set of fields--antifields, which will be labeled by relaxed world-sheet picture numbers. 
As a result, we find that {\it additional propagating degrees of freedom} appear in loop amplitudes of superstring field theory. 
Our second motivation is to clarify these. 

Since the set of fields--antifields is enlarged, the resultant Batalin-Vilkovisky master action $S_{\sf bv}$ can take some different forms. 
In particular, one cannot obtain unique $S_{\sf bv}$ by just relaxing the ghost number constraint unlike well-established cases. 
It would be a natural consequence from relaxing {\it geometry-inspired} constraints on the fields--antifields. 
However, at the same time, it would imply the existence of a larger class of consistent master actions for superstring field theory, which is our third motivation. 
Interestingly, this unrestricted $S_{\sf bv}$ and its behaviour under canonical transformations recall the WZW-like formulation \cite{Berkovits:1995ab, Berkovits:2004xh, Matsunaga:2014wpa, Erler:2015uoa, Kunitomo:2015usa, Matsunaga:2015kra, Goto:2015pqv, GK, Matsunaga:2016zsu, Erler}. 

\sub{Organization of the article}

In section 2, we give an analysis of the \textit{large} gauge symmetry of superstring field theory based on the \textit{small} Hilbert space. 
On the basis of the Batalin-Vilkovisky formalism, we derive corresponding master action, which takes a slightly different form from usual one. 
We study this enlarged master action and show that canonical transformations bridge the gap from the well-established analysis. 
In section 3, we clarify the hidden gauge reducibility and the underlying \textit{large} gauge structure of superstring field theory based on the \text{small} Hilbert space. 
In section 4, we see how these additional fields--antifields appear in the canonical form of the master action. 
Then, we give master actions for interacting theories based on the set of fields--antifields without geometry-inspired restrictions. 
We end with remarks on the relation between the hidden gauge reducibility, the large class of the master action, and the general WZW-like formulation. 

\section{Large gauge symmetry of small theory} 

The gauge structure of superstring field theory is infinitely reducible. 
We know that the $g$-th gauge parameter $\lambda _{1-g}$ has its gauge invariance $\delta \lambda _{1-g} = Q \, \lambda _{-g}$\, and it is preserved under the $(g+1)$-st gauge transformation $\lambda _{-g} \rightarrow \lambda _{-g} + \delta \lambda _{-g}$ with $\delta \lambda _{-g} = Q \lambda _{-(g+1)} $\,. 
However, there is an implicit assumption: 
As well as a dynamical string field $\Psi$ and its gauge variation $\delta \Psi$, all gauge parameter fields must belong to the small Hilbert space, $\{ \lambda _{1-g} \} _{g} \subset \cH _{\beta \gamma }$. 
Note that using the zero mode $\eta \equiv \eta _{0}$ of $\eta (z)$-current of the $\xi \eta \phi $-system, one can express this restriction as 
\begin{align*}
\lambda _{1-g} \in \cH _{\beta \gamma } 
\hspace{5mm} \Longleftrightarrow \hspace{5mm} 
\eta \, \lambda _{1-g} = 0 \, , \hspace{5mm} (g \in \mathbb{N}) \, .
\end{align*} 
We often call this restriction onto the small Hilbert space $\cH _{\beta \gamma }$ as the $\eta $-constraint. 
If and only if we impose this constraint on not only the string field but also all gauge parameter fields, the theory has this type of the gauge reducibility. 

The above gauge reducibility structure is drastically changed by relaxing the $\eta$-constraint on the gauge transformations for the gauge transformations, keeping on the action and its gauge invariance. 
To see it explicitly at the level of the Batalin-Vilkovisky master action, it is useful to switch from the small BPZ inner product $\slla A , B \srra $ to the large BPZ inner product $\langle A,B \rangle $: 
\begin{align}
\label{kinetic} 
S [\Psi ] = \frac{1}{2} \la \xi \, \Psi , \, Q \, \Psi \ra \equiv \frac{1}{2} \lla \Psi , \, Q \, \Psi \rra \, . 
\end{align}
Here, $\xi $ is a homotopy operator for $\eta$, namely, $\eta \, \xi + \xi \, \eta = 1$ in the large Hilbert space $\cH _{\xi \eta \phi }$. 
Since $2 \delta S = - \langle \delta \Psi , (Q \xi - \xi Q) \Psi \rangle $\,, we find the gauge invariance $\delta \Psi = Q \, \lambda$ provided that $\eta \, \Psi = 0$ and $\eta ( \delta \Psi ) = 0$\,. 
It implies that $\lambda $ must satisfy $\eta \, \lambda = Q \, \omega $ with some state $\omega$\,: 
To be gauge invariant, the constraint $\eta \, \lambda = 0$ (and $\eta \, \lambda _{-g} = 0$ for $g \in \mathbb{N}$) is sufficient but not necessary. 
In this section, we see what happens if we relax $\eta \lambda = 0$ under $\eta \Psi = 0$ and $\eta (\delta \Psi ) = 0$. 

\vspace{1mm} 

Note that one can apply the analysis which we present below to every types of superstring fields $\Psi \in \cH _{\beta \gamma }$. 
The mismatch of their Grassmann parities based on world-sheet ghost numbers is resolved by using the grading based on the appropriately suspended degrees. 
For example, see \cite{Gaberdiel:1997ia, Erler:2014eba} or appendix A. 
However, for simplicity, one can regard $g,p$ of $\Psi _{g,p}$ as the ghost and picture number labels of open superstring fields in the rest of this section.

\sub{Unrestricted gauge parameter} 

Let us consider the kinetic term of \cite{Witten:1986qs, Jurco:2013qra, Erler:2013xta, Erler:2014eba, Sen:2015uaa, Konopka:2016grr, Erler:2016ybs, Ohmori:2017wtx}, namely the free theory (\ref{kinetic}). 
Now, we impose the $\eta$-constraints on the dynamical string field $\Psi $ and its gauge variation $\delta \Psi $ only, 
\begin{align}
\label{free constraint}
\eta \, \Psi = 0 \, , \hspace{5mm} \eta \, ( \delta \Psi ) = 0 \, .
\end{align}
Because of these constraints, any assignment of $\xi$ in the action (\ref{kinetic}) can be permissible, and we can rewrite the variation of (\ref{kinetic}) as $\delta S  = - \la \delta \Psi , \, Q \, \xi \Psi \ra $\,. 
We write $\Psi _{1,-1} \equiv \Psi$\,. 
The $p$-label of $\Psi _{g,p}$ denotes its world-sheet picture number.\footnote{For type II theory, we should write it as $\Psi _{g,p,\tilde{p}}$ and consider the left- and right-moving picture numbers} 
In this set up, the action is invariant under the gauge transformation 
\begin{align}
\label{free gauge transformation}
\delta \Psi _{1,-1} = Q \, \lambda _{0,-1} \, . 
\end{align}
Now, a gauge parameter string field $\lambda _{0,-1} \equiv \lambda$ has to keep (\ref{free constraint}), but there is no other restrictions. 
We consider to enlarge the state space of $\lambda _{0,-1}$ from $\cH _{\beta \gamma }$ to $\cH _{\beta } \oplus Q \, \cH _{\xi \eta \phi }$\,. 
Clearly, it keeps the form of the gauge transformation (\ref{free gauge transformation}) and the constraint $\delta \Psi _{1,-1}\in \cH _{\beta \gamma }$ because {\it all enlarged components of} $\lambda _{0,-1}$ {\it have no new-contributions into} (\ref{free gauge transformation}).\footnote{This is our implicit assumption in section 2. 
If we remove it, the space $\cH _{\beta } \oplus Q \, \cH _{\xi \eta \phi }$ does not give the largest extension. 
Then, we can further enlarge it to ${\rm Ker} [ Q \eta ]$, in which the same story goes by replacing $\cH _{\beta } \oplus Q \, \cH _{\xi \eta \phi }$ with ${\rm Ker}[Q \eta ]$. We discuss it again in section 3. The author would like to thank Ted Erler.} 
Then, because of $\eta (Q \lambda _{0,-1})= 0$\,, there exists a gauge parameter $\lambda _{0,-2}$ such that 
\begin{subequations} 
\begin{align}
\label{weaken constraint}
\eta \, \lambda _{0,-1} + Q \, \lambda _{0,-2} = 0 \, . 
\end{align} 
It gives a weaker constraint on $\lambda _{0,-1}$ rather than $\eta \, \lambda _{0,-1} = 0$\,. 
Here, $\lambda _{0,-2}$ is an auxiliary gauge parameter string field which belongs to $\cH _{\beta \gamma } \oplus Q \, \cH _{\xi \eta \phi }$. 
It is not the end of story: 
This weaken constraint (\ref{weaken constraint}) implies $\eta (Q\lambda _{0,-2} ) = 0$\,, which provides the constraint also on $\lambda _{0,-2}$\,. 
Again, using an auxiliary gauge parameter $\lambda _{0,-3}$, we have to impose $\eta \, \lambda _{0,-2} + Q \, \lambda _{0,-3} = 0$\,.  
As a result, we find that for the gauge transformation (\ref{free gauge transformation}) keeping the $\eta $-constraint (\ref{free constraint}), there exists a family of infinite number of the auxiliary gauge parameter string fields $\{ \lambda _{0,-p} \} _{p }$ satisfying 
\begin{align} 
\label{weaken constraint b}
\eta \, \lambda _{0,-p} + Q \, \lambda _{0,-(p+1)} = 0 \,, \hspace{5mm} ( p \in \mathbb{N} ) \, . 
\end{align} 
\end{subequations} 
We therefore find that in the theory based on (\ref{free constraint}), while the dynamical string field and its gauge variation must belong to the small Hilbert space, its gauge parameter and auxiliary fields can protrude from the small Hilbert space as long as they are BRST exact: 
\begin{align*}
\Psi _{1,-1} \in \cH _{\beta \gamma } \, , \hspace{5mm} 
\lambda _{0,-p} \in \cH _{\beta \gamma } \oplus Q \, \cH _{\xi \eta \phi } \, . 
\end{align*} 

\sub{Gauge reducibility with constraints} 

The relations (\ref{weaken constraint}) and (\ref{weaken constraint b}) will give the first class constraints on the set of fields--antifields \cite{Batalin:1992mk}. 
On the basis of the gauge transformation (\ref{free gauge transformation}) and gauge parameter fields living in $\cH _{\beta \gamma } \oplus Q \, \cH _{\xi \eta \phi }$, we study the gauge reducibility of superstring field theory. 
Note that there is no positive-picture gauge parameter fields $\lambda _{0,+p}$\, for $p \in \mathbb{N}$ in the above set up.\footnote{One may be possible to introduce additional gauge parameter field $\lambda _{0,0}$ and consider a similar tower of auxiliary fields satisfying $\eta \, \Psi _{0,0} + Q \, \Psi _{0,-1} = 0$; see section 3. 
It would correspond to switch the roles of $Q$ and $\eta$ in the following analysis. 
Since $Q$- and $\eta$-complexes are exact in the large Hilbert space $\cH _{\xi \eta \phi }$, one could introduce a homotopy operator for $Q$ and consider the switched version in the completely same way.}
We consider the gauge transformations $\delta _{1}$ for the gauge transformation (\ref{free gauge transformation}), which must preserve (\ref{free constraint}), 
\begin{align*}
\delta _{1} ( \delta \Psi _{1,-1}  ) = 0 \, , 
\hspace{5mm} 
\delta _{1} ( \eta \, \lambda _{0,-p} + Q \, \lambda _{0,-(p+1)} ) = 0 \,, \hspace{5mm} ( p \in \mathbb{N} ) \, . 
\end{align*}
In other words, for $p \in \mathbb{N}$\,, we want to specify $\delta _{1} \lambda _{0,-p}$ such that  
$Q ( \delta _{1} \lambda _{0,-1} ) = 0$ and $\eta \, ( \delta _{1} \lambda _{0,-p} ) + Q \, ( \delta _{1} \lambda _{0,-(p+1)} ) = 0$\,. 
We find the first higher gauge transformations, 
\begin{align*} 
\delta _{1} \lambda _{0,-1}  = Q \, \lambda _{-1,-1} \, , 
\hspace{5mm}  
\delta _{1} \lambda _{0,-(p+1)}  = \eta \, \lambda _{-1,-p} + Q \, \lambda _{-1,-(p+1)} \, .
\end{align*}
Next, we consider the gauge transformations $\delta _{2}$ preserving these, $\delta _{2} ( \delta _{1} \lambda _{0,-1} ) = 
Q \, ( \delta _{2} \lambda _{-1,-1} ) = 0$ and $\eta \, ( \delta _{2} \lambda _{-1,-p} ) + Q \, ( \delta _{2} \lambda _{-1,-(p+1)} ) = 0$ for $p \in \mathbb{N}$\,. 
We find that again, they are given by $\delta _{2} \lambda _{-1,-1} = Q \, \lambda _{-2,-1}$ and $\delta _{2} \lambda _{-1,-(p+1)} = \eta \, \lambda _{-2,-p} + Q \, \lambda _{-2,-(p+1)}$ for $p \in \mathbb{N}$. 
Likewise, for $p \in \mathbb{N}$\,, we obtain the $g$-th gauge transformations 
\begin{align} 
\label{free higher gauge transformations}
\delta _{g} \lambda _{-g,-1}  = Q \, \lambda _{-(g+1),-1} \, , 
\hspace{5mm}  
\delta _{g} \lambda _{-g,-(p+1)}  = \eta \, \lambda _{-(g+1),-p} + Q \, \lambda _{-(g+1),-(p+1)} \, , 
\end{align}
which preserves the $(g-1)$-th gauge transformations 
\begin{align*}
\delta _{g} \big( \delta _{g-1} \lambda _{-(g-1),-1} \big) & = Q \big( \delta _{g} \lambda _{-g,-1} \big) = 0 \, ,
\\
\delta _{g} \big( \delta _{g-1} \lambda _{-(g-1),-1-p} \big) & = \eta \big( \delta _{g} \lambda _{-g,-p} \big) + Q \big( \delta _{g} \lambda _{-g,-1-p} \big) = 0 \, .
\end{align*} 
Hence, as well as the theory based on the restriction $\lambda _{1-g} \in \cH _{\beta \gamma }$, our unrestricted theory based on $\delta \Psi \in \cH _{\beta \gamma }$ has an infinitely reducible gauge structure. 
However, in our case, the gauge hierarchy takes a more enlarged form. 
As we will see, this change of the gauge reducibility yields the different spectrum of fields--antifields and Batalin-Vilkovisky master action.

\subsection{Solving the Batalin-Vilkovisky master equation} 

The above analysis of the gauge reducibility tells us that in addition to the string field $\Psi $ and the (first) gauge parameter $\lambda $, an infinite tower of higher gauge parameters $\{ \lambda _{1-g,-p} \} _{g,p>0}$ appears in the theory. 
Hence, we introduce the set of fields--antifields as follows. 
\begin{align*}
{\rm Fields} \, \, : & \, \hspace{5.5mm}
\Psi _{1,-1} \, , \hspace{7mm} 
\{ \Psi _{0, -1-p} \} _{p \in \mathbb{N}} \, , \hspace{6.5mm} 
\{ \Psi _{-1, -1-p} \} _{p \in \mathbb{N}} \, , \hspace{4.5mm} 
\dots \, , \hspace{4mm} 
\{ \Psi _{1-g, -1-p} \} _{p \in \mathbb{N}} \, , \hspace{4.5mm} 
\dots \, 
\\ 
{\rm Antifields} \, \, : & \, \hspace{3mm} 
\underbrace{(\Psi _{1,-1})^{\ast } }_{\Phi ^{\ast }_{1,0}} \, , \hspace{2mm} 
\underbrace{ \{ (\Psi _{0, -1-p})^{\ast } \} _{p \in \mathbb{N}} }_{ \{ \Phi ^{\ast }_{2,p} \} _{p \in \mathbb{N}} } \, , \hspace{2mm} 
\underbrace{ \{ (\Psi _{-1, -1-p})^{\ast } \} _{p \in \mathbb{N}} }_{ \{ \Phi ^{\ast }_{3,p} \} _{p \in \mathbb{N}} } \, , \hspace{2mm} 
\dots \, , \hspace{2mm} 
\underbrace{ \{ (\Psi _{1-g, -1-p})^{\ast } \} _{p \in \mathbb{N}} }_{ \{ \Phi ^{\ast }_{1+g,p} \} _{p \in \mathbb{N}} } \, , \hspace{2mm} 
\dots \, 
\end{align*}
This is a non-minimal set of fields--antifields and there exist the first class constraints corresponding to (\ref{weaken constraint}) and (\ref{weaken constraint b}) \cite{Batalin:1992mk, Berkovits:2012np}. 
For Neveu-Schwarz (NS) open string fields, the field $\Psi _{1-g,-1-p} \equiv A_{g} \cZ _{1-g,-1-p}$ consists of a set of space-time fields $A_{g}$ whose space-time ghost number is $g$ and a set of CFT basis $\cZ _{1-g,-1-p}$ whose world-sheet ghost and picture numbers are $1-g$ and $-1-p$\,. 
When we use the large BPZ inner product to construct the master action,\footnote{One may wonder if it changes the properties of the master action which are valid when we the small BPZ inner product, such as $K_{\sf bv} = K(\psi , Q \psi)$. 
But it is not the case. 
The use of the large BPZ inner product provides just a framework to describe the modification or enlargement, and it is mainly caused by the BV spectrum and constraints on fields--antifields, which we will see in section 3.2. 
See also appendix A. } the corresponding antifields $(\Psi _{1-g ,-1-p})^{\ast }$ consists of a set of space-time fields $A_{-(g+1)}$ whose space-time ghost number is $-(g+1)$ and a set of CFT basis $\cZ _{1+g,p}$ whose world-sheet ghost and picture numbers are $1+g$ and $p$\,, namely, $(\Psi _{1-g , -1-p})^{\ast } \equiv A_{-(g+1)} \, \cZ _{1+g, p}$\,. 
Thus, for $g , p \geq 0$\,, we write $\Phi ^{\ast }_{1+g , p}$ for the antifield of $\Psi _{1-g,-1-p}$ as follows 
\begin{align*}
\Phi ^{\ast }_{1+g, p} \equiv ( \Psi _{1-g , -1-p} )^{\ast } \,.
\end{align*}
The Grassmann parities of fields and antifields are different: $(-)^{\Psi } = (-)^{\Phi +1}$. 
In this paper, $g$ of $\Psi _{1-g,-1-p}$ or $\Phi ^{\ast }_{1+g,p}$ denotes the $g$-th reducibility, and $p$ of that indicates the $p$-decreasing from the natural picture number of considering string fields. 
Because of (\ref{weaken constraint}), the subset of fields $\{ \Psi _{1,-1} , \Psi _{0,-p} \} _{p > 0}$ must satisfy the constraint equations 
\begin{align}
\label{higher constraints}
\eta \, \Psi _{1,-1} = 0 \, , \hspace{5mm} 
\eta \, \Psi _{0,-p} + Q \, \Psi _{0,-1-p} = 0 \, . 
\end{align}
There is no constraint on the other fields or antifields. 
We write $\cG$ for the gauge generator of the theory, namely, $\cG \equiv Q$ in this case and $\cG \equiv \{ Q , V_{g,n } \} _{g,n}$ for the interacting theory discussed in section 3. 
We therefore consider the following BV spectrum with constraints (\ref{higher constraints}) 
\begin{align}
\label{primary constraints}
\Psi _{1,-1} \in \cH _{\beta \gamma } \, , \hspace{3mm} 
\{ \Psi _{0,-p} \} _{p>0} \subset \cH _{\beta \gamma } \oplus \cG \,\cH _{\xi \eta \phi }  \, , \hspace{3mm} 
\{ \Psi _{-g,-p} \,, \Phi _{g,p-1}^{\ast } \} _{g,p > 0} \subset \cH _{\xi \eta \phi } \, . 
\end{align} 
We construct a BV master action based on (\ref{primary constraints}) and (\ref{higher constraints}) and see its properties in the rest of this section. 
As we will see, the constraints (\ref{higher constraints}), or simply $\eta \, \Psi _{1,-1} = Q \, \eta \, \Psi _{0,-1} = 0$, work well and play an important role in the BV master equation. 

\sub{Antifield number expansion}

We derive the master action $S_{\sf bv} [ \Psi , \Phi ^{\ast } ]$ on the basis of the antifield expansion, 
\begin{align}
\label{afn expansion}
S_{\sf bv} [ \Psi , \Phi ^{\ast } ] = S^{(0)} [\Psi ] + \sum_{a=1}^{\infty } S^{(a)} [ \Psi , \Phi _{\ast } ] \, . 
\end{align}
Here, $S^{(a)}$ denotes the antifield number $a$ part of the master action $S_{\sf bv}$. 
The antifield number is additive and assigned to antifields only: the $a$-th antifield $\Phi _{a,p}^{\ast }$ has antifield number $a$, for which we write ${\rm afn}[\Phi _{a,p}^{\ast }] = a$. 
For simplicity, we define the antifield number of the field $\Psi _{g,p}$ by $0$, namely ${\rm afn}[\Psi _{g,p} ] = 0$. 
Every functions $F=F[\Psi , \Phi ^{\ast }]$ of fields--antifields $\{ \Psi , \Phi ^{\ast } \}$ have the antifield number which is equivalent to the sum of inputs antifield numbers. 
Therefore, note that the derivative with respect to the antifield $\Phi _{a,p}^{\ast }$ must decrease antifield number $a$. 
The BPZ inner product does not have the antifield number: ${\rm afn} [ \langle A , B \rangle ] = {\rm afn}[A] + {\rm afn} [ B]$. 

As the initial condition of the master action, the antifield number $0$ part $S^{(0)}$ is given by the original action $S [\Psi ]$ itself, 
\begin{align*}
S^{(0)} [ \Psi  ] \equiv S [ \Psi ] = \frac{1}{2} \la \xi \, \Psi _{1,-1} , \, Q \, \Psi _{1,-1} \ra \, . 
\end{align*}
Let us consider the antifield expansion of the master equation $\{ S_{\sf bv} , S_{\sf bv} \} = \sum_{g=0}^{\infty } \{ S_{\sf bv} , S_{\sf bv} \} |^{(g)}$, where $\{ \, ,\, \}$ denotes the antibracket; see appendix A. 
The antifield number $g$ part is given by 
\begin{align*}
\Big{\{ } \, S_{\sf bv} [\Psi , \Phi ^{\ast } ] \, , \, S_{\sf bv} [\Psi , \Phi ^{\ast } ] \, \Big{\} } \Big{|}^{(g)} 
= \sum_{p=0}^{\infty } \LA \frac{\partial _{r} S^{(g)}}{\partial \Psi _{1-g,-1-p}} , \, \frac{\partial _{r} S^{(1+g)}}{\partial \Phi ^{\ast }_{1+g,p}}  \RA \, . 
\end{align*}
First, we specify the antifield number $1$ part $S^{(1)}$ of $S_{\sf bv}$, which has to satisfy the antifield number $0$ part of the master equation with $S^{(0)}$: 
\begin{align*}
\Big{\{ } \, S^{(0)} + S^{(1)} \, , \, S^{(0)} + S^{(1)} \, \Big{\} } \Big{|}^{(0)} = 
\sum_{p} \LA \frac{\partial _{r} S^{(0)}}{\partial \Psi _{1,-1}} , \, \frac{\partial _{r} S^{(1)}}{\partial \Phi ^{\ast }_{0,p}}  \RA = 0 \, . 
\end{align*}
Note that because of (\ref{primary constraints})\,, using real parameters $a,b \in \mathbb{R}$ satisfying $a+ b \not= 0$\,, we obtain  
\begin{align*}
\frac{\partial _{r} S^{(0)}}{\partial \Psi _{1,-1}} = \frac{1}{a+b} \big( a \, \xi \, Q - b \, Q \, \xi \big) ( \eta \, \xi \, \Psi _{1,-1} ) 
= \frac{1}{a+b} \big( a \, \xi \, Q - b \, Q \, \xi \big) \Psi _{1,-1} \, . 
\end{align*}
To solve the equation $\{ S^{(0)} + S^{(1)} , \, S^{(0)} + S^{(1)} \} |^{(0)} = 0$\,, we set 
\begin{align*}
\frac{\partial _{l} S^{(1)}}{\partial \Phi ^{\ast }_{1,0}} = Q \, \Psi _{0,-1} 
\overset{(\ref{primary constraints})}{=} \eta \xi (Q \, \Psi _{0,-1}) \, . 
\end{align*} 
One may think this is a natural and unique choice. 
However, note that the constraint relation (\ref{primary constraints}) is crucial to satisfy the master equation using this right derivative, unlike usual cases. 
We find an antifield number $1$ part of the solution, 
\begin{subequations} 
\begin{align}
\label{free small S^1}
S^{(1)} [\Psi , \Phi ^{\ast } ] & = \la \Phi ^{\ast }_{1,0} , \, Q \, \Psi _{0,-1} \ra \, . 
\end{align}
In this case, unlike well-established analysis based on the restriction $\{ \Psi _{g}, (\Psi _{g})^{\ast } \} _{g} \subset \cH _{\beta \gamma }$, the projector $\eta \xi $ does not work as the identity on fields, which leads another expression. 
This $S^{(1)}$ provides the right derivative with respect to the ghost string field $\Psi _{0,-1}$\,, 
\begin{align*}
\frac{\partial _{r} S^{(1)} }{\partial \Psi _{0,-1}}  = Q \, \Phi _{1,0}^{\ast } \, . 
\end{align*} 
In this expression of $S^{(1)}$\,, clearly, other additional ghost derivatives vanish. 
Note however that the recursive relation $\eta \, \Psi _{0,-p} + Q \, \Psi _{0,-1-p} = 0$ on $\Psi _{0,-p} = (\eta \, \xi + \xi \, \eta ) \Psi _{0,-p}$ implies 
\begin{align*}
Q \, \Psi _{0,-1} 
& = Q \, \big( \eta \, \xi \, \Psi _{0,-1} \big) - Q \, X \, \Psi _{0,-2} 
\no 
& = Q \, \eta \, \xi \, \big( \Psi _{0,-1} - X \, \Psi _{0,-2} \big) + Q \, X^{2} \, \Psi _{0,-3}  
\no & \hspace{20mm}
\vdots 
\no 
& = Q \, ( \eta \, \xi \, \Psi _{0,-1}) + Q \sum _{p=1}^{\infty } (-)^{p} X^{p} \, (\eta \, \xi \,\Psi _{0,-1-p} ) \, . 
\end{align*}
Here, $X \equiv Q \, \xi + \xi \, Q$ changes the picture number. 
We obtain alternative expression of the left derivative with respect to $\Phi _{1,0}^{\ast }$\,. 
In this alternative expression, a projector $\eta \xi $ is inserted in front of the states, which resolves the ambiguity of $\xi$-assignments. 
Note that because of the constraints (\ref{higher constraints}), all ghost fields $Q \, \Psi _{0,-p}$ $(p \in \mathbb{N})$ have this expression. 
Hence, $S^{(1)}[\Psi , \Phi ^{\ast }]$ potentially includes all auxiliary fields $\Psi _{0,-1-p}$\,, and we can rewrite (\ref{free small S^1}) as follows, 
\begin{align}
\label{free S^1}
S^{(1)} [\Psi , \Phi ^{\ast } ] 
= \la \Phi ^{\ast }_{1,0} , \, Q \, ( \eta \xi \Psi _{0,-1} ) \ra 
+ \sum_{p=1}^{\infty } (-)^{p} \la \Phi ^{\ast }_{1,0} , \, Q \, X^{p} ( \eta \xi  \Psi _{0,-1-p} ) \ra \, . 
\end{align}
\end{subequations} 
Interestingly, each term individually vanishes in the master equation. 
Thus, if one prefers, one could set different coefficients for each term at this level. 
It suggests that it may be possible to use $X$ explicitly for solving the master equation, which we discuss later. 
This expression of $S^{(1)}$ provides the right derivatives with respect to each ghost string field $\{ \Psi _{0,-p} \} _{p \in \mathbb{N}}$\,, 
\begin{align} 
\label{1st ghost derivatives} 
\frac{\partial _{r} S^{(1)}}{\partial \Psi _{0,-1-p}} 
& = - (-)^{p} \xi \, Q \, X^{p} ( \eta \, \Phi _{1,0}^{\ast } ) \, . 
\end{align}
Note that $\eta $ and $\xi$ are inserted into the $\Psi _{0,-1}$-derivative in this expression. 

Before solving the next order equation, let us consider about $X$-including terms of (\ref{free S^1})\,. 
When we impose the rigid small-space constraint $\Psi _{0,-1} = \eta \xi (\Psi _{0,-1})$ on the ghost field, these additional ghost terms should vanish and (\ref{free S^1}) reduces to (\ref{free small S^1}) with $\Psi _{0,-1} \in \cH _{\beta \gamma }$\,. 
If ghost field $\Psi _{0,-1}$ is exactly small, the consistent $\Psi _{0,-1}$-ghost derivative should be a $\xi$-exact state in the large BPZ inner product. 
It implies that the antifield $\Phi _{1,0}^{\ast }$ satisfies  
\begin{align*}
Q \, \Phi _{1,0}^{\ast } = \xi \eta \, ( Q \, \Phi _{1,0}^{\ast } ) 
\end{align*}
in $S^{(1)}$. 
This constraint on $\Phi _{1,0}^{\ast }$\,, the antifield for the string field $\Psi _{1,-1}$\,, yields  
\begin{align*}
X \, \eta \, Q \, \Phi _{1,0}^{\ast } = 0 \, , 
\end{align*}
which kills (\ref{1st ghost derivatives}) and additional terms appearing in (\ref{free S^1})\,.

One can also see the consistency of two expression (\ref{free small S^1}) and (\ref{free S^1}) via direct computations of the next order BV master equation. 
We consider the antifield number $2$ part $S^{(2)}$, which has to satisfy the antifield number $1$ part of the master equation, 
\begin{align*}
\frac{1}{2}\Big{\{ } S_{\sf bv} [ \Psi , \Phi ^{\ast } ] \, , \, S_{\sf bv} [ \Psi , \Phi ^{\ast } ] \Big{\} } \Big{|}^{(1)} = 
\sum_{p} 
\LA \frac{\partial _{r} S^{(1)}}{\partial \Psi _{-1,-1-p}} , \, \frac{\partial _{l} S^{(2)}}{\partial \Phi ^{\ast }_{2,p}} \RA = 0 \, . 
\end{align*}
From the ghost derivatives of $S^{(1)}$\,, we find that left derivatives should be given by 
\begin{align*}
\frac{\partial _{l} S^{(2)}}{\partial \Phi ^{\ast }_{2,0}} = Q \, \Psi _{-1,-1} \, , 
\hspace{8mm} 
\frac{\partial _{l} S^{(2)}}{\partial \Phi ^{\ast }_{2,1+p}} = \eta \, \Psi _{-1,-1-p} + Q \, \Psi _{-1,-2-p} \, . 
\end{align*}
While we quickly find that these satisfy the master equation if we use (\ref{free small S^1})\,, 
on the basis of (\ref{free S^1})\,, we obtain the following pieces of the master equation, 
\begin{align*}
\LA \frac{\partial _{r} S^{(1)}}{\partial \Psi _{0,-1}} , \, \frac{\partial _{r} S^{(2)}}{\partial \Phi ^{\ast }_{2,0}} \RA 
& = \La \xi \, Q \, \eta \, \Phi _{1,0}^{\ast } , \, Q \, \Psi _{-1,-1} \Ra 
= \La \eta \, Q \, X \, \Phi _{1,0}^{\ast } , \, \Psi _{-1,-1} \Ra \, , 
\\ 
\sum_{p=1}^{\infty } \LA \frac{\partial _{r} S^{(1)}}{\partial \Psi _{0,-1-p}} , \, \frac{\partial _{r} S^{(2)}}{\partial \Phi ^{\ast }_{2,p}}  \RA 
& =
\sum_{p=1}^{\infty } (-)^{p} \La Q \, \eta \, X^{p} \Phi ^{\ast }_{1,0} , \, 
\Psi _{-1,-p} + X \, \Psi _{-1,-1-p} \Ra \, \, .
\end{align*}
By summing up all $p$-labels, these satisfy the master equation 
\begin{align*}
\frac{1}{2} \big{\{ } S_{\sf bv} , \, S_{\sf bv} \big{\} } \big{|}^{(1)} = \sum_{p=0}^{\infty } (-)^{p} \La ( Q \, \eta + \eta \, Q ) \, X^{p} \, \Phi _{1,0}^{\ast } , \, \Psi _{-1,-1-p} \Ra = 0 \, . 
\end{align*}
Hence two expression are consistent, and at this step, we obtain the following solution 
\begin{align*}
S^{(2)} [ \Psi , \Phi ^{\ast } ] = \la \Phi ^{\ast }_{2,0} , \, Q \, \Psi _{-1,-1} \ra + \sum_{p=1}^{\infty } \la \Phi ^{\ast }_{2,p} , \, \eta \, \Psi _{-1,-p} + Q \, \Psi _{-1,-1-p} \ra  \, . 
\end{align*}
Its ghost derivatives are given by 
\begin{align*}
\frac{\partial _{r} S^{(2)}}{\partial \Psi _{-1,-1-p} }  =  \eta \, \Phi ^{\ast }_{2,1+p} + Q \, \Phi ^{\ast }_{2,p}  \, . 
\end{align*}
To satisfy the next order master equation, $\{ S_{\sf bv} , S_{\sf bv} \} |^{(2)} = 0$\,, we have to set the antighost derivatives of $S^{(3)}$ as follows 
\begin{align*}
\frac{\partial _{r} S^{(3)}}{\partial \Phi ^{\ast }_{3,0} } = Q \, \Psi _{-2,-1} \, ,
\hspace{5mm}  
\frac{\partial _{r} S^{(3)}}{\partial \Phi ^{\ast }_{3,p} }  = \eta \, \Psi _{-2,-p} + Q \, \Psi _{-2,-1-p} \, , 
\end{align*} 
which gives the antifield number $3$ part of the solution $S^{(3)}$ in the similar form as $S^{(2)}$. 
Likewise, we find that the antifield number $g$ part of $S_{\sf bv}$ is given by 
\begin{align*}
S^{(g)} [ \Psi , \Phi ^{\ast } ] = \la \Phi ^{\ast }_{g,0} , \, Q \, \Psi _{1-g,-1} \ra + \sum_{p=1}^{\infty } \la \Phi ^{\ast }_{g,p} , \, \eta \, \Psi _{1-g,-p} + Q \, \Psi _{1-g,-1-p} \ra  \, . 
\end{align*}
Note that this type of the antifield number $g$ part $S^{(g)}$ works as Lagrange-multiplier-like ghost--antighost term fixing the higher gauge symmetries (\ref{free higher gauge transformations}).

\subsection{Master action and BRST transformations} 

To see how constraints (\ref{primary constraints}) work in the master equation, we introduce a set of Lagrange multipliers $\cL = \{ \cL _{0,1}, \, \cL _{1,p} \} _{p \in \mathbb{N} }$\,. 
Note that $\cL _{g,p}$ has space-time ghost number $-g$, world-sheet ghost number $g$\,, and picture number $p$\,. 
By summing up all antifield number $S^{(g)}$, we get 
\begin{subequations} 
\begin{align}
\label{with Lmp}
S_{\sf bv} [ \Psi , \Phi ^{\ast } ; \cL  ] & = 
\frac{1}{2} \la \xi \Psi _{1,-1} , \, Q \, \Psi _{1,-1} \ra 
+ \sum_{g=1}^{\infty } \la \Phi ^{\ast }_{1+(g-1),0} , \, Q \, \Psi _{1-g , -1} \ra 
\no & \hspace{12mm}
+ \sum_{g = 1}^{\infty } \sum_{p=1}^{\infty } \la \Phi ^{\ast }_{1+g,p} , \, \eta \, \Psi _{-g,-p} + Q \, \Psi _{-g,-(p+1)} \ra \,
\no & \hspace{8mm} 
+ \,\, \la \cL _{0,1} , \, \eta \, \Psi _{1,-1} \ra 
+ \,\, \sum_{p=1}^{\infty } \, \la \cL _{1,p} , \, \eta \, \Psi _{0,-p} + Q \, \Psi _{0 , -(p+1)} \ra \, . 
\end{align}
After integrating out the Lagrange multipliers $\cL$ of (\ref{with Lmp}), we obtain the BV master action, 
\begin{align}
\label{free master action} 
S_{\sf bv} [ \Psi , \Phi ^{\ast } ] = \int \cD [\cL ] \, S_{\sf bv} [ \Psi , \Phi ^{\ast } ; \cL ] \, .
\end{align} 
\end{subequations} 
We derived the master action using the antifield number expansion. 
But off course, because of the free theory, one can apply the BRST formalism as its gauge fixing procedure and find $S_{\sf bv}$ by the guess from it as \cite{Torii:2011zz, Kroyter:2012ni, Torii:2012nj}. 

\sub{Master equation}

Let $\delta _{\rm BV} \Psi $ and $\delta _{\rm BV} \Phi ^{\ast }$ be BV-BRST transformations of fields and antifields respectively. 
We check that our master action $S_{\sf bv} [\Psi , \Phi ^{\ast } ]$ satisfies the master equation 
\begin{align}
\label{free master eq} 
\Big{\{ } \, S_{\sf bv} [ \Psi , \Phi ^{\ast } ] \, , \, S_{\sf bv} [ \Psi , \Phi ^{\ast } ] \, \Big{\} } 
= 0 \, . 
\end{align} 
Then, the Batalin-Vilkovisky formalism implies that BRST transformations are given by 
\begin{align*}
\delta _{\rm BV} \Psi _{1-g,-1-p} & = \frac{\partial _{l} S_{\sf bv}[\Psi , \Phi ^{\ast }]}{\partial \Phi ^{\ast }_{1+g, p} }
= \Big{\{ } \, \Psi _{1-g,-1-p} \, , \, S_{\sf bv} [\Psi , \Phi ^{\ast } ] \, \Big{\} } \, , 
\\ 
\delta _{\rm BV} \Phi ^{\ast }_{1+g,p} & = \frac{\partial _{l} S_{\sf bv}[\Psi , \Phi ^{\ast }]}{\partial \Psi _{1-g, -1-p} } 
= \Big{\{ } \, \Phi ^{\ast }_{1+g,p} \, , \, S_{\sf bv} [\Psi , \Phi ^{\ast } ] \, \Big{\} } \, . 
\end{align*}

We write $S_{\sf bv} [ \cL ] \equiv S_{\sf bv} [ \Psi , \Phi ^{\ast } ; \cL ]$ for brevity. 
In the previous section, we found that the derivatives with respect to the first pair of field--antifield $\{ \Psi _{1,-1} , \, \Phi ^{\ast }_{1,0} \}$ are given by 
\begin{align*} 
\frac{\partial _{r} S_{\sf bv} [\cL ] }{\partial \Psi _{1,-1}}  = \frac{a \, \xi \, Q - b \, Q \, \xi }{a + b} \Psi _{1,-1} 
+ \eta \, \cL _{0,1} \, , \hspace{5mm} 
\frac{\partial _{r} S_{\sf bv} [\cL ] }{\partial \Phi ^{\ast }_{1,0}}  = Q \, \Psi _{0,-1} \, . 
\end{align*}
where $a,b \in \mathbb{R}$ are real parameters satisfying $a + b \not= 0$\,. 
For the primary ghosts $\{ \Psi _{1-g,-1} \} _{g}$ and auxiliary ghosts $\{ \Psi _{1-g,-1-p} \} _{g,p}$\,, we obtained their right derivatives as 
\begin{align*} 
\frac{\partial _{r} S_{\sf bv} [\cL ] }{\partial \Psi _{1-g,-1-p}}  & = \eta \, \Phi ^{\ast }_{g,1+p} + Q \, \Phi ^{\ast }_{g,p} 
 + \delta _{1,g} \big( \eta \, \cL _{g,p+1} + Q \, \cL _{g,p} \big) \, , 
\end{align*}
for given $g \in \mathbb{N}$ and $p\in \{ 0 \} \cup \mathbb{N}$\,. 
For the primary antighosts $\{ \Phi _{1+g,0}^{\ast } \} _{g}$ and auxiliary antighosts $\{ \Phi ^{\ast }_{1+g,p} \} _{g,p}$ with any fixed $g, p \in \mathbb{N}$\,, their left derivatives are given by 
\begin{align*} 
\frac{\partial _{l} S_{\sf bv} [\cL ] }{\partial \Phi ^{\ast }_{1+g,0}} = Q \, \Psi _{-g,-1} \, , 
\hspace{5mm} 
\frac{\partial _{l} S_{\sf bv} [\cL ] }{\partial \Phi ^{\ast }_{1+g,p}} & = \eta \, \Psi _{-g,-p} + Q \, \Psi _{-g,-(p+1)} \, . 
\end{align*}
Using these, we find that up to the terms including Lagrange multipliers, at each level of the $g$-label, the master equation holds after the sum over the $p$-label: 
\begin{align*}
\frac{1}{2} \big{\{ } S_{\sf bv} [ \cL ] , \, S_{\sf BV} [\cL ] \big{\} } 
& 
=\sum_{g =0}^{\infty } \LA \frac{\partial _{r} S_{\sf BV}  [ \cL ] }{\partial \Psi _{1-g, -1} } , \, \frac{\partial _{l} S_{\sf BV}  [ \cL ] }{\partial \Phi ^{\ast }_{1+g,0}} \RA 
+ \sum_{g =1}^{\infty } \sum_{p=1}^{\infty } \LA \frac{\partial _{r} S_{\sf BV}  [ \cL ] }{\partial \Psi _{1-g, -1 - p} } , \, \frac{\partial _{l} S_{\sf BV}  [ \cL ] }{\partial \Phi ^{\ast }_{1+g,p}} \RA 
\no & 
= \LA \frac{\partial _{r} S_{\sf bv} [\cL ] }{\partial \Psi _{1, -1} } , \, \frac{\partial _{l} S_{\sf bv} [\cL ] }{\partial \Phi ^{\ast }_{1,0}} \RA 
+ \sum_{g =0}^{\infty } \sum_{p=1}^{\infty } \La \big( \Phi ^{\ast }_{g,p} + \cL _{1,p} \big) , \, ( Q \, \eta + \eta \,Q ) \Psi _{-g,-p} \Ra \, .
\end{align*} 
We would like to emphasise that the second term vanishes without requiring higher constraint equations on fields, which may reduce (\ref{higher constraints}) and give $\Psi _{0,p<-1} \in \cH _{\xi \eta \phi }$\,. 
As a result, we obtain 
\begin{align*}
\frac{1}{2} \big{\{ } S_{\sf bv} [ \cL ]  , \, S_{\sf bv} [ \cL ] \big{\} } 
=  \la Q \, \Psi _{0,-1} , \, \xi \, Q \, \Psi _{1,-1} \ra 
+ \la \cL _{0,1} , \, \eta \, Q \, \Psi _{0,-1} \ra  \, , 
\end{align*} 
which clearly reduces to zero when $\eta \, \Psi _{1,-1}= 0$ and $\eta \, Q \, \Psi _{0,-1} = 0$ hold. 
While we introduced $\cL _{0,1}$ to impose the $\eta $-constraint on the string field $\Psi _{1,-1} \in \cH _{\beta \gamma }$\,, in the master equation, it also works to impose the $\eta $-constraint on the gauge variation $\delta _{\rm BV} \Psi _{1,-1} \in \cH _{\beta \gamma }$\,. 
It completes a proof that the action (\ref{free master action}) satisfies the master equation (\ref{free master eq}) and that the master action $S_{\sf bv} [\Psi , \Phi ^{\ast } ]$ is invariant under the BRST transformations 
\begin{align*} 
\delta _{\rm BV} \Psi _{1-g,-1}  & = Q \, \Psi _{1-g,-1} \, ,
\hspace{5.5mm} 
\delta _{\rm BV} \Psi _{-g,-1-p}  = \eta \, \Psi _{-g-1,-p} + Q \, \Psi _{-g-1,-1-p} \, ,
\\
\delta _{\rm BV} \Phi ^{\ast }_{1,0}  & = Q \, \xi \, \Psi _{1,-1} \, ,
\hspace{12mm} 
\delta _{\rm BV} \Phi ^{\ast }_{1+g,p}  = \eta \, \Phi ^{\ast }_{g,p+1} + Q \, \Phi ^{\ast }_{g,p} \, . 
\end{align*}

\subsection{Reduction to the small master action} 

We will see that the master action (\ref{free master action}) reduces to that of exactly small theory. 
Roughly, by integrating out the additional antifields $\{ \Phi _{g,p} \} _{p>0}$ of (\ref{free master action}), all ghost fields $\{ \Psi _{-g.-p} \} _{g,p}$ are restricted on the subspace $\Sigma $ satisfying the constraint equations $\eta \, \Psi _{-g,1-p} + Q \, \Psi _{-g,-p} = 0$\,. 
Then, $Q \, \Psi _{-g,-1} = \eta \xi ( Q \, \Psi _{-g,-1} )$ because of $\eta \, ( Q \, \Psi _{-g,-p} ) = 0$\,, and we find 
\begin{align*}
S_{\sf bv} [ \Psi _{-}, \Phi _{+}^{\ast } ] \big{|}_{\Sigma } & = 
\frac{1}{2} \la \xi \Psi _{1,-1} , \, Q \, \Psi _{1,-1} \ra 
+ \sum_{g=1}^{\infty } \la \Phi ^{\ast }_{g,0} , \, \eta \xi ( Q \, \Psi _{1-g , -1} ) \ra \, 
\no 
& = \frac{1}{2} \La \, \xi \, ( \Psi _{-} + \eta \, \Phi _{+}^{\ast } ) \, , \, Q \, ( \Psi _{-} + \eta \, \Phi _{+}^{\ast } ) \, \Ra \, ,  
\end{align*}
where $\Psi _{-} \equiv \Psi _{1,-1} + \sum_{g=1}^{\infty } \Psi _{1-g,-1}$ and $\Phi _{+}^{\ast } \equiv \Phi _{1,0}^{\ast } + \sum_{g=1}^{\infty } \Phi _{1+g,0}^{\ast }$\,. 
This is the small BV master action based on the large BPZ inner product, exactly small fields $\Psi _{-} \in \cH _{\beta \gamma }$, and unrestricted antifields $\Phi _{+}^{\ast } \in \cH _{\xi \eta \phi }$. 
Hence, by identifying $\eta \, \Phi _{+}^{\ast }$ with the antifield $\Psi _{\textrm{s}}^{\ast }$ of the exactly small theory, $\Psi _{\textrm{s}}^{\ast } \cong \eta \, \Phi _{+}^{\ast }$, or by imposing constraints $\xi \, \Phi _{+}^{\ast } = 0$, which is equivalent to restrict the minimal set of fields--antifields onto $\cH _{\beta \gamma }$, it reduces to the small BV master action based on the small BPZ inner product and $\{ \Psi _{-} , \Psi _{\textrm{s}}^{\ast } \} \subset \cH _{\beta \gamma }$. 
See appendix B for the small theory. 
In terms of the Batalin-Vilkovisky formalism, it implies that there exists a (partially) gauge fixing fermion which reduces (\ref{free master action}) to the small master action, which we explain.  
 
\sub{Gauge fixing fermion}

Using two types of trivial pairs of BV fields $\{ C_{g,1} , N_{g+1,1} \} _{g}$ and $\{ A_{g,p} , L_{g+1,p} \} _{g,p}$, we add an auxiliary term, a trivial solution of the master equation 
\begin{align*}
S_{\sf aux} [ L , N ; A^{\ast } , C^{\ast } ] = \sum _{g=-\infty }^{\infty } \la C_{2-g,-2}^{\ast } , \, N_{g,1} \ra 
+ \sum_{g=1}^{\infty } \sum_{p=1}^{\infty } \la A_{1-g,-1-p}^{\ast } , \, L _{1+g,p} \ra \, , 
\end{align*}
into the master action (\ref{free master action}). 
Here, $C_{2-g,-2}^{\ast }$ and $A_{1-g,-1-p}^{\ast }$ are antifields for $C_{g,1}$ and $A_{1+g,p}$ respectively. 
We also introduce BV fields $\{ \Psi _{1+g,-1}^{\dagger } \} _{g}$, which will be identified with the antifields of the small master action. 
Let us consider the following gauge fixing fermion $\Gamma [\psi ] = \Gamma [ \Psi , A ,C ; \Psi ^{\dagger } ]$ consisting of this non-minimal set of fields, 
\begin{align*}
\Gamma [\psi ] 
& = \sum_{g=1}^{\infty } \Big[ \la C_{-g,1} , \, \eta \, \Psi ^{\dagger }_{1+g,-1} \ra 
+ \la  \xi \, \Psi ^{\dagger }_{1+g,-1} + \eta \, C_{g,1} , \, \Psi _{1-g,-1} \ra 
+\sum_{p=1}^{\infty } \la A_{1+g,p} , \, \Psi _{1-g,-1-p} \ra 
\Big] \, .
\end{align*} 
It gives the following ghost field derivatives 
\begin{align*}
& \hspace{12mm} 
\frac{\partial \, \Gamma [\psi ] }{\partial \Psi _{1-g,-1}} = \xi \Psi _{1+g,-1}^{\dagger } + \eta \, C_{g,1} \, ,
\hspace{5mm}  
\frac{\partial \, \Gamma [\psi ]}{\partial \Psi _{-g,-p}} = A_{2+g,p-1} \, ,
\\ & 
\frac{\partial \, \Gamma [\psi ]}{\partial C_{g>0,1}} = \eta \, \Psi _{1-g,-1} \, ,
\hspace{5mm}
\frac{\partial \, \Gamma [\psi ]}{\partial C_{g<0,1}} = \eta \, \Psi _{1+g,-1}^{\dagger } \, ,
\hspace{5mm} 
\frac{\partial \, \Gamma [\psi ]}{\partial A_{1+g,p}} = \Psi _{1-g,-1-p} \, . 
\end{align*}
On this gauge fixing fermion, we have $\Phi ^{\ast } \equiv \partial _{\Psi }\Gamma $ and $S [ \psi ; \psi ^{\ast } ] |_{\Gamma }= S [ \psi ; \partial _{\psi } \Gamma ]$. 
Therefore, by integrating out $\{ L , N \}$\,, we obtain the small BV master action 
\begin{align*}
S_{\sf bv} [\Psi _{-} , \Psi _{+}^{\dagger } ] & = \int \cD [ L ] \, \cD [ N ] \, \Big( S_{\sf bv} [\Psi , \partial _{\Psi } \Gamma ] + S_{\sf aux} [ L , N ; \partial _{A} \Gamma , \partial _{C} \Gamma ] \Big) \,
\no 
& = \frac{1}{2} \La \, \xi \, (\Psi _{-} + \Psi _{+}^{\dagger } ) \, , \, Q \, ( \Psi _{-} + \Psi _{+}^{\dagger } ) \, \Ra \, , 
\end{align*}
where $\Psi _{-} \equiv \sum _{g=0}^{\infty } \Psi _{1-g,-1}$ and $\Psi _{+}^{\dagger } \equiv \sum _{g=0}^{\infty } \Psi _{2+g,-1}^{\dagger }$\,. 
Note that $\eta \Psi _{1-g,-1} = \eta \Psi _{2+g,-1}^{\dagger } = 0$ and $\{ \Psi _{g,p} \} _{p<1}=0$ because of $N$- and $L$-integrations. 

\vspace{2mm} 

Therefore, the well-established BV master action based on the geometry-inspired constraints $\Psi _{-} \in \cH _{\beta \gamma }$ kills the large gauge symmetries of higher ghost fields and is equivalent to a partially gauge-fixed version of the gauge theory without restrictions.

\subsection{Canonical transformations} 

It is known that the Batalin-Vilkovisky master action is unique up to adding trivial pairs and canonical transformations if it is proper. 
In this section, we discuss three important types of canonical transformations. 
In particular, we show that there exist a canonical transformation which rotates only higher ghost-fields--antifields and transforms the master action (\ref{free master action}) into   
\begin{align}
\label{canonical master action}
S_{\sf bv} [ \Psi , \Phi ^{\ast } ] = \frac{1}{2} \la \xi \, \Psi _{1,-1} , \, Q \, \Psi _{1,-1} \ra 
+ \sum_{g=0}^{\infty } \sum_{p=0}^{\infty } \la \Phi _{1+g,p}^{\ast } , \, Q \, \Psi _{1-g,-1-p} \ra \, , 
\end{align}
where we used $\Phi _{1,p>0} = 0$ for brevity and $\{ \Psi _{1,-1} , \Psi _{-g,-1-p} , \Phi _{1,0}^{\ast } , \Phi _{2+g,p}^{\ast } \} _{g,p}$ is the set of fields--antifields. 
Although it has the same form as the master action based on the geometry-inspired constraints $\{ \Psi _{g} , (\Psi _{g})^{\ast } \} _{g} \subset \cH _{\beta \gamma }$, it includes additional propagating ghost--antighost fields.\footnote{Thus, one could take a short-cut to the reduction presented in the previous section by finding an appropriate gauge-fixed basis killing additional fields--antifields of (\ref{canonical master action}).} 
We call $(\ref{canonical master action})$ as the canonical form, and $(\ref{free master action})$ as the large form. 

\sub{On the explicit $X$-insertions} 

Recall that because of the constraints (\ref{higher constraints}), all ghost fields $Q \, \Psi _{0,-p}$ $(p \in \mathbb{N})$ have another expression, and $Q \, X^{p} \Psi _{0,-1-p}$ satisfies the antifield number $0$ part of the master equation for any $p \in \mathbb{N}$\,. 
Hence, for example, we could start from 
\begin{align*}
S^{(1)} [\Psi , \Phi ^{\ast } ] 
& = \la \Phi ^{\ast }_{1,0} , \, Q \,  \Psi _{0,-1}  \ra 
+ \sum_{p=1}^{\infty } (-)^{p} \la \Phi ^{\ast }_{1,0} , \, Q \, X^{p} \Psi _{0,-1-p} \ra \, . 
\end{align*}
If one prefers, one could use different coefficients for each term. 
Using (\ref{primary constraints}), this $S^{(1)}$ provides the right derivatives with respect to ghost string fields $\{ \Psi _{0,-p} \} _{p \in \mathbb{N}}$\,, 
\begin{align} 
\label{X ghosts}
\frac{\partial _{r} S^{(1)}}{\partial \Psi _{0,-1-p}} 
& = (-)^{p} X^{p} \, Q \, \Phi _{1,0}^{\ast } 
+ (-)^{p+1} X^{p+1} \eta \, \Phi _{1,0}^{\ast } \, . 
\end{align}
This $S^{(1)}$ also satisfies $\{ S_{\sf bv} , S_{\sf bv} \} |^{(1)} = 0$ via the same mechanism as we found, 
\begin{align*}
\sum_{p=0}^{\infty } \LA \frac{\partial _{r} S^{(1)}}{\partial \Psi _{0,-1-p}} , \, \frac{\partial _{r} S^{(2)}}{\partial \Phi ^{\ast }_{2,p}}  \RA 
= \sum_{p=0}^{\infty } (-)^{p} \La ( Q \, \eta + \eta \, Q ) \, X^{p+1} \Phi ^{\ast }_{1,0} , \, \widehat{\Psi }_{-1,-1-p} \Ra = 0\, , 
\end{align*}
and it leads the same type of the master action as (\ref{free master action}). 
These master actions will be related to each other via a canonical transformation. 
For fixed $g,p \geq 0$\,, we defined  
\begin{subequations} 
\begin{align}
\widehat{\Psi }_{1-g,-1-p} & \equiv \Psi _{1-g,-1-p} + \sum_{q=1}^{\infty } (-)^{q} X^{q} \Psi _{1-g,-1-p-q} \, , 
\end{align}
which gives the field relation between new and old pairs of the fields--antifields. 
In this notation, we can rewrite the above $S^{(1)}$ and its antighost derivative as 
\begin{align*}
S^{(1)} = \la \widehat{\Phi }_{1,0}^{\ast } , \, Q \, \widehat{\Psi }_{0,-1} \ra \, , 
\hspace{8mm} 
\frac{\partial _{l} S^{(1)}}{\partial \widehat{\Phi }^{\ast }_{1,0}} & = Q \, \widehat{\Psi }_{0,-1}  \, . 
\end{align*} 
Here, we write $\widehat{\Phi }_{1+g,p}^{\ast }$ for the antifield corresponding to the field $\widehat{\Psi }_{1-g,-1-p}$\,. 
Then, from it ghost derivatives or (\ref{X ghosts}), we find  
\begin{align*}
\frac{\partial _{l} S^{(2)}}{\partial \widehat{\Phi }^{\ast }_{2,0}} = Q \, \widehat{\Psi }_{-1,-1} \, , 
\hspace{8mm} 
\frac{\partial _{l} S^{(2)}}{\partial \widehat{\Phi }^{\ast }_{2,1+p}} = \eta \, \widehat{\Psi }_{-1,-1-p} + Q \, \widehat{\Psi }_{-1,-2-p} \, . 
\end{align*}
These give the same type of the master action. 
Note that via the generating function of this canonical transformation, antifields $\Phi _{1+g,p}^{\ast }$ and $\widehat{\Phi }_{1+g,p}^{\ast }$ are related by 
\begin{align}
\Phi _{1+g,p}^{\ast } = \widehat{\Phi }^{\ast }_{1+g,p} + \sum_{q=1}^{p} (-)^{q} X^{q} \widehat{\Phi }^{\ast }_{1+g,p-q} \, . 
\end{align}
\end{subequations} 
Therefore, ambiguity of the form of $S_{\sf bv}$ coming up from using explicit $X$-insertions can be absorbed by canonical transformations. 

\sub{Switching transformation} 

Interestingly, there exist canonical transformations switching the roles of $\eta $ and $Q$. 
We consider the following generating function $\cR [ \Psi , \widehat{\Phi }^{\ast } ]$ of the canonical transformation 
\begin{align*}
\cR [ \Psi , \widehat{\Phi }^{\ast } ] & = 
\sum_{g=0}^{\infty } 
\la \widehat{\Phi }^{\ast }_{1+g,0} , \Psi _{1-g,-1} \ra 
+ \sum_{g=1}^{\infty } \sum_{p=1}^{\infty }
\la \widehat{\Phi }_{1+g,p}^{\ast } , \, \eta \, \xi \, \Psi _{1-g,-1-p} - \xi \, Q \, \Psi _{1-g,-2-p} \ra \, .  
\end{align*} 
Apparently, it leaves pairs of field-antifield labeled by $p=0$ invariant. 
The new fields $\{ \widehat{\Psi } , \widehat{\Phi }^{\ast } \}$ and old fields $\{ \Psi , \Phi ^{\ast } \}$ are related by 
\begin{align*}
\widehat{\Psi }_{-g,-1-p} \equiv \frac{\partial _{l} \cR [ \Psi , \widehat{\Phi }^{\ast }] }{\partial \widehat{\Phi }_{2+g,p}^{\ast }} \, , 
\hspace{8mm}  
\Phi _{2+g,p}^{\ast } \equiv \frac{\partial _{r} \cR [ \Psi , \widehat{\Phi }^{\ast } ] }{\partial \Psi _{-g,-1-p} } \, . 
\end{align*} 
Since the first ghost fields satisfy (\ref{higher constraints}), we find $\widehat{\Psi }_{0,-1-p} = (\eta \xi + \xi \eta ) \Psi _{0,-1-p}$. 
Thus, $\cR [ \Psi , \widehat{\Phi }^{\ast } ]$ generates identity transformation not only for the $p=0$ subset $\{ \Psi _{1-g,-1} , \Phi _{1+g,0}^{\ast } \} _{g}$, but also the first ghost-fields--antifields $\{ \Psi _{0,-1-p} , \Phi _{1,p}^{\ast } \} _{p}$. 
For generic $g,p \geq 1$, it gives  
\begin{align*}
\widehat{\Psi }_{-g,-1-p} 
= \eta \, \xi \, \Psi _{-g,-1-p} - \xi \, Q \, \Psi _{-g,-2-p} \, , 
\hspace{5mm}  
\Phi _{2+g,p}^{\ast } 
=  \xi \, \eta \, \widehat{\Phi }_{2+g,p}^{\ast } - Q \, \xi \, \widehat{\Phi }_{2+g, p-1}^{\ast }  \, . 
\end{align*}
Using these, for the higher ghost-fields--antifields, one can quickly find 
\begin{align*}
\la \Phi _{1+g,1+p}^{\ast } , Q \, \Psi _{-g,-2-p} \ra 
& = \la \xi \, \eta \, \widehat{\Phi }_{1+g,1+p}^{\ast } , \, Q \, \Psi _{-g,-2-p} \ra 
\no & 
= \la \, \widehat{\Phi }_{1+g,1+p}^{\ast } , \, \eta \, \big( \eta \, \xi \, \Psi _{-g,-1-p} - \xi \, Q \, \Psi _{-g,-2-p} \big) \ra \, . 
\end{align*}
By summing up all $g,p \geq 0$, we can transform the higher ghost-fields--antifields terms of the master action (\ref{canonical master action}) into 
\begin{align*}
\sum_{g=1}^{\infty } \sum_{p=1}^{\infty }
\la \Phi _{1+g,p}^{\ast } , Q \, \Psi _{-g,-1-p} \ra & = 
- \sum_{g=1}^{\infty } \sum_{p=1}^{\infty }
\la \widehat{\Phi }_{1+g,p}^{\ast } , \eta \, \widehat{\Psi }_{-g,-p} \ra  \, . 
\end{align*}
The minus sign of the right hand side would be natural from the point of view of the $\eta $-$Q$ switching relation appearing in WZW-like superstring field theory. 
If one prefer, one can absorbe this sign by redefining the fields--antifields or by appropriate canonical transformations.\footnote{For example, one can perform $\cI _{g,p} [ \Psi , \widehat{\Phi }^{\ast }] = i \langle \widehat{\Phi }_{1+g,p}^{\ast } , \Psi _{1-g,-1-p} \rangle $ or more trivial transformations.} 
While we introduced the canonical transformation leaving the $p=0$ subset $\{ \Psi _{1-g,-1} , \Phi _{1+g,0}^{\ast } \} _{g}$\,, one can consider canonical transformations switching all pair of the fields--antifields similarly. 

\subsubsection*{The canonical form and the large form}

After the above canonical transformation switching $\eta$- and $Q$-terms, we consider to take back the $\cR [\Psi , \widehat{\Phi }^{\ast }]$-transformed master action via the following canonical transformation 
\begin{align*}
\cW [ \Psi , \widehat{\Phi }^{\ast } ] & =  
\la \widehat{\Phi }^{\ast }_{1,0} , \Psi _{1,-1} \ra +
\sum_{g=0}^{\infty }  
\sum_{p=0}^{\infty }  
\la \widehat{\Phi }^{\ast }_{2+g,p} , \Psi _{-g,-1-p} \ra 
+ \sum_{g=1}^{\infty } \sum_{p=1}^{\infty }
\la \widehat{\Phi }_{2+g,p}^{\ast } , \, 
\xi \, Q \, \Psi _{-g,-2-p} \ra \, .  
\end{align*}
Two minimal sets of fields--antifields $\{ \Psi , \Phi \}$ and $\{ \widehat{\Psi } , \widehat{\Phi }^{\ast } \}$ are related by 
\begin{align*}
\widehat{\Psi }_{-g,-1-p} \equiv \frac{\partial _{l} \cW [ \Psi , \widehat{\Phi }^{\ast }] }{\partial \widehat{\Phi }_{2+g,p}^{\ast }} \, , 
\hspace{5mm}  
\Phi _{2+g,p}^{\ast } \equiv  \frac{\partial _{r} \cW [ \Psi , \widehat{\Phi }^{\ast } ] }{\partial \Psi _{-g,-1-p} } \, . 
\end{align*}
By construction, $\cW [\Psi , \widehat{\Phi }^{\ast } ]$ acts as the identity on the $p=0$ subset $\{ \Psi _{1-g,-1} , \Phi _{1+g,0}^{\ast } \} _{g}$ and on the first ghost-fields--antifields $\{ \Psi _{0,-1-p} , \Phi _{2,p}^{\ast } \} _{p}$\,. 
For other $g,p >0$, it generates 
\begin{align*}
\widehat{\Psi }_{-g,-1-p} = \Psi _{-g,-1-p} + \xi \, Q \, \Psi _{-g,-2-p} \, , 
\hspace{5mm}  
\Phi _{2+g,p}^{\ast } = \widehat{\Phi }_{2+g,p}^{\ast } + Q \, \xi \, \widehat{\Phi }_{2+g, p-1}^{\ast }  \, . 
\end{align*}
Therefore, via $\cW [\Psi , \widehat{\Phi }^{\ast }]$, the higher ghost-fields--antifields terms are transformed as  
\begin{align*}
\la \widehat{\Phi }_{1+g,p}^{\ast } , \, \eta \, \widehat{\Psi }_{-g,-p} \ra 
& = \la \widehat{\Phi }_{1+g,p}^{\ast } , \, \eta \, \Psi _{-g,-p} + \eta \xi Q \, \Psi _{-g,-1-p} \ra 
\no 
& = \la \widehat{\Phi }_{1+g,p}^{\ast } , \, Q \, \Psi _{-g,-1-p} \ra 
+ \la  \widehat{\Phi }_{1+g,p}^{\ast } , \, \eta \, \Psi _{-g,-p} + \xi \, Q \, \eta \, \Psi _{-g,-1-p} \ra \, . 
\end{align*}
By summing up all $g,p \geq 0$\,, we obtain 
\begin{align*}
\sum_{g=1}^{\infty } \sum_{p=1}^{\infty }
 \la \widehat{\Phi }_{1+g,p}^{\ast } , \eta \, \widehat{\Psi }_{-g,-p} \ra
= \sum_{g=1}^{\infty } \sum_{p=1}^{\infty } \Big[ 
\la \Phi _{1+g,p}^{\ast } , Q \, \Psi _{-g,-1-p} \ra 
+ \la \Phi _{1+g,p}^{\ast } , \eta \, \Psi _{-g,-p} \ra \Big] \, . 
\end{align*}
Hence, there exists a canonical transformation between the large form of the master action (\ref{free master action}) and the canonical form of the master action (\ref{canonical master action}). 
Note that this canonical transformation does not change the dynamical string field, its antifield, and string fields of the first ghost-fields--antifields: 
It rotates string fields of the additional ghost-fields--antifields only. 

The use of the large Hilbert space $\cH _{\xi \eta \phi }$ enable us to consider various forms of $S_{\sf bv}$ and canonical transformations drastically changing $S_{\sf bv}$. 
It would make quantization of large superstring field theory based on the WZW-like formulation highly complicated problem.\footnote{See \cite{Matsunaga:2018hlh} for the BV formalism in the large Hilbert space: Several classical BV master actions were obtained. } 

\section{Hidden gauge reducibility}

In section 2, the hidden gauge symmetries arising from $\delta \Psi _{1,-1} = Q \, \lambda _{0,-1}$ were revealed by enlarging the space of the gauge parameter $\lambda _{0,-1}$ from $\cH _{\beta \gamma }$ to $\cH _{\beta \gamma } \oplus Q \, \cH _{\xi \eta \phi }$ (or ${\rm Ker} [Q \eta ]$\footnote{While $\lambda _{0,-1} \in (\eta \oplus Q)\cH _{\xi \eta \phi }$ has no new-contributions into $\delta \Psi _{1,-1} = Q \, \lambda _{0,-1}$, however, $\lambda _{0,-1} \in {\rm Ker}[Q \eta ]$ does. 
We consider ${\rm Ker} \big[ \eta \big] \cup {\rm Ker} \big[ Q \big]$ in this paper, but one can consider $ {\rm Ker} \big[ Q \eta \big] $ instead of it in the same way. }). 
What is the origin of these \textit{large} gauge symmetries in the \textit{small} theory?---we clarify it in this section. 
Recall that the variation of the action (\ref{kinetic}) is given by 
\begin{align*}
\delta S [ \Psi ] 
= - \la \delta \Psi _{1,-1} , \, Q \, \xi \, \Psi _{1,-1}  \ra 
+ \frac{1}{2} \la \delta \Psi _{1,-1} , \, X \Psi _{1,-1}  \ra \, ,
\end{align*}
where $X = \xi \, Q + Q \, \xi$\,. 
The second term always vanishes because of $\delta \Psi \in \cH _{\beta \gamma }$\,. 
It gives the on-shell condition $Q \, \Psi _{1,-1} = 0 $, which is invariant under $\delta \Psi _{1,-1} = Q \, \lambda _{0,-1}$ with $\eta \, (Q \, \lambda _{0,-1})= 0$. 
However, the condition $\eta ( \delta \Psi ) = \eta  ( Q \lambda _{0,-1} ) = 0$ implies that there exists $\omega _{0,0}$ such that $Q \lambda _{0,-1} = \eta \, \omega _{0,0}$. 
It provides another expression of this gauge transformation $\delta \Psi = \eta \, \omega _{0,0}$ with $Q ( \eta \, \omega _{0,0}) = 0$. 
If we permit to use this ambiguous expression, it yields more enlarged gauge hierarchy, which we first explain in this section. 
Alternatively, we can identify $\lambda _{0,-1} \equiv \eta \, \Lambda _{-1,0}$ by using a \textit{large} gauge parameter $\Lambda _{-1,0} \in \cH _{\xi \eta \phi }$. 
As we explain, it gives unambiguous expression of the hidden gauge reducibility and clarifies the origin of the large gauge symmetry. 

\subsection{Hidden gauge reducibility with the first class constraints} 

We write $2 \, \mu _{0,-1} \equiv \lambda _{0,-1}$ and $2 \, \mu _{0,0} \equiv \omega _{0,0}$ for brevity. 
Note that $\delta S [\Psi ] = 0$ holds if 
\begin{align*}
\delta \Psi \in {\rm Ker} [\eta ] \cap {\rm Ker}[Q] = ( \eta \, \cH _{\xi \eta \phi } ) \cap ( Q \, \cH _{\xi \eta \phi }) \, . 
\end{align*}
We therefore find that the action is invariant under the gauge transformations 
\begin{align}
\label{re-expression}
\delta \Psi _{1,-1} = Q \, \mu _{0,-1} + \eta \, \mu _{0,0} \, . 
\end{align}
These two gauge parameters $\{ \mu _{0,-k} \} _{k=0,1}$ belong to the kernel of $\eta $ or $Q$, 
\begin{align} 
\label{ker}
\mu _{0,-1} , \, \mu _{0,0} \in  {\rm Ker} \big[ \eta \big] \cup {\rm Ker} \big[ Q \big] \, ,
\end{align}
and they are not independent $Q \, \mu _{0,-1} = \eta \, \mu _{0,0}$\,. 
The state $Q \, \mu _{0,-1}$ or the state $\eta \, \mu _{0,0}$ lives in the subspace ${\rm Ker} [\eta ] \cap {\rm Ker} [Q]$ because of $\mu _{0,-1} , \mu _{0,0} \in  {\rm Ker} \big[ Q \big] \cup  {\rm Ker} \big[ \eta \big]$. 

To see the relation of $\mu _{0,-1}$ and $\mu _{0,0}$ explicitly, it may be helpful to recall that ${\rm Ker} [\eta ] \cup {\rm Ker} [Q] \subset {\rm Ker}[Q \eta ]$ holds. 
Let us consider two states $V_{0,0} , \,V_{0,-1}  \in {\rm Ker}[Q\eta ]/ ({\rm Im}[Q] \cup {\rm Im}[\eta ])$. 
We consider $\mu _{0,-1}, \mu_{0,0} \in {\rm Ker} \big[ Q \, \eta \big]$ instead of (\ref{ker}), which reduce to (\ref{ker}) by setting $V_{0,0} = V_{0,-1} = 0$ below. 
We also introduce three states $\widehat{\mu }_{-1,-1}, \widehat{\mu }_{-1,0}, \widehat{\mu }_{-1,1} \in \cH _{\xi \eta \phi }$ living in the large Hilbert space. 
Since these $\mu _{0,-1}$ and $\mu _{0,0}$ live in ${\rm Ker}[Q\eta ]$, we can write them as follows, 
\begin{align*}
\mu _{0,-1} & = V_{0,-1} 
+ \eta \, \widehat{\mu }_{-1,0} + Q \, \widehat{\mu }_{-1,-1} \, , 
\no  
\mu _{0,0} & = V_{0,0} 
- Q \, \widehat{\mu }_{-1,0} - \eta \, \widehat{\mu }_{-1,1} \, .
\end{align*}
By construction, $V_{0,-1}$ and $V_{0,0}$ must satisfy $Q \, V _{0,-1} = \eta \, V_{0,0}$. 
We thus write $V_{1,-1} \equiv \eta \, V_{0,0} = Q \, V_{0,-1}$, which gives $V_{0,-1} = Q^{-1} V_{1,-1}$ and $V_{0,0} = \xi \, V_{1,-1}$ in the above expression. 
Note that $V_{1,-1} \in {\rm Ker} [\eta ] \cap {\rm Ker} [Q]$. 
Using these $V_{1,-1}$ and $\widehat{\mu }_{-1,0}$, we find  
\begin{subequations} 
\begin{align}
\label{dep of mu1}
Q \, \mu _{0,-1} = V_{1,-1} + Q \, \eta \, \widehat{\mu }_{-1,0} = \eta \, \mu _{0,0} \, . 
\end{align}

The hidden gauge reducibility of superstring field theory based on the small Hilbert space is essentially provided by the {\it large} gauge variation $\delta \widehat{\mu }_{-1,0} = Q \, \widehat{\mu }_{-2,0} + \eta \, \widehat{\mu }_{-2,1}$ preserving (\ref{dep of mu1}), where $\widehat{\mu }_{-1,0}$, $\widehat{\mu }_{-2,0}$, and $\widehat{\mu }_{-2,1}$ all live in the large Hilbert space $\cH _{\xi \eta \phi }$. 

\sub{Auxiliary gauge parameters}

The (first) gauge parameters $\mu _{0,-1}$ and $\mu _{0,0}$ live in $ {\rm Ker} \big[ \eta \big] \cup {\rm Ker} \big[ Q \big]$. 
It implies that the state $\eta \, \mu _{0,-1}$ lives in ${\rm Ker}[Q] = Q \, \cH _{\xi \eta \phi }$ and the state $Q \, \mu _{0,0}$ lives in ${\rm Ker} [\eta ] = \cH _{\beta \gamma } \equiv \eta \, \cH _{\xi \eta \phi } $\,. 
We find auxiliary gauge parameters $\mu _{0,-2}$ and $\mu _{0,-1}$ such that 
\begin{align*}
\eta \, \mu _{0,-1} \in {\rm Ker}[Q] \,  
\hspace{3mm} & \Longleftrightarrow \hspace{3mm} 
\eta \, \mu _{0,-1} + Q \, \mu _{0,-2} = 0 \, , 
\no 
Q \, \mu _{0,0} \in {\rm Ker} [\eta ] \, 
\hspace{3mm} & \Longleftrightarrow \hspace{3mm} 
Q \, \mu _{0,0} + \eta \, \mu _{0,1} = 0 \, .
\end{align*}
However, note that these $\mu _{0,-2}$ and $\mu _{0,-1}$ must satisfy $\eta \, Q \mu _{0-2} = 0$ and $\eta \, Q \mu _{0,-1} = 0$. 
Hence, these auxiliary parameters $\mu _{0,-2}$ and $\mu _{0,1}$ live in the same subspace as $\mu _{0,-1}$ and $\mu _{0,0}$, 
\begin{align*}
\mu _{0,-2} \, , \hspace{1mm} \mu _{0,1} \in 
{\rm Ker} \big[ \eta \big] \cup {\rm Ker} \big[ Q \big]  \, .  
\end{align*}
Likewise, we find a set of auxiliary gauge parameters $\{ \mu _{0,p} \} _{p \not= -1,0} \subset {\rm Ker} \big[ \eta \big] \cup {\rm Ker} \big[ Q \big]$ such that 
\begin{align}
\label{dep of mu2}
\eta \, \mu _{0,p+1} + Q \, \mu _{0,p} = 0 \, , \hspace{5mm} (p \not= -1) \, . 
\end{align}
\end{subequations}  
The (first) gauge parameters $\{ \mu _{0,-1} , \, \mu _{0,0} \}$ and all auxiliary gauge parameters $\{ \mu _{0,p} \} _{p \not = -1,0}$ belong to the kernel of $\eta $ or $Q$, 
\begin{align*}
\big{\{ } \mu _{0,p} \big{\} }_{p \in \mathbb{Z} } \subset {\rm Ker} \big[ \eta \big] \cup {\rm Ker} \big[ Q \big] \, . 
\end{align*}
Note that there gauge parameters are dependent each other through (\ref{dep of mu1}) and (\ref{dep of mu2}). 

\sub{Higher gauge transformations}

The above gauge transformation (\ref{re-expression}) is completely equivalent to $\delta \Psi _{1,-1} = Q \, \lambda _{0,-1}$ since it is just a redefinition of the gauge parameters. 
However, as we see, all higher gauge parameters (and those of auxiliary gauge parameters) appearing in its gauge reducibility are independent each other unlike $\{ \mu _{0,p} \} _{p\in \mathbb{Z}} \subset {\rm Ker} \big[ \eta \big] \cup {\rm Ker} \big[ Q \big]$, which would be an interesting point. 

We find the gauge transformations preserving the first gauge transformation (\ref{re-expression}) 
\begin{align*}
\delta \mu _{0,-1} & = Q \, \mu _{-1,-1} + \eta \, \mu _{-1, 0} \, , 
\no 
\delta \mu _{0,0} & = Q \, \mu _{-1,0} + \eta \, \mu _{-1,1} \, , 
\end{align*}
and the gauge transformations of auxiliary gauge parameter fields 
\begin{align*}
\delta \mu _{0,p} = Q \, \mu _{-1,p} + \eta \, \mu _{-1,p+1} \, , \hspace{5mm} (p \not= -1,0) \, ,
\end{align*}
where all gauge parameters $\{ \mu _{-1,p} \} _{p\in \mathbb{Z}}$ belong to the large Hilbert space: $\{ \mu _{-1,p} \} _{p \in \mathbb{Z} } \subset \cH _{\xi \eta \phi } $\,. 
The new ingredients are higher gauge parameters $\mu _{-1,p}$ labeled by positive $p$, and the $p$-label runs over all integer numbers. 
They are invariant under the third gauge transformations 
\begin{align*}
\delta \mu _{-1,-1} & = Q \, \mu _{-2,-1} + \eta \, \mu _{-2, 0} \, , 
\no 
\delta \mu _{-1,0} & = Q \, \mu _{-2,0} + \eta \, \mu _{-2,1} \, , 
\no
\delta \mu _{-1,1} & = Q \, \mu _{-2,1} + \eta \, \mu _{-2,2} \, , 
\end{align*}
and the second gauge transformations of auxiliary gauge parameters 
\begin{align*}
\delta \mu _{-1,p} = Q \, \mu _{-2,p} + \eta \, \mu _{-2,p+1} \, ,\hspace{5mm} (p \not= -1,0,1) \, .  
\end{align*}
Likewise, we find the $(g+1)$-st gauge transformations preserving $g$-th gauge transformations 
\begin{subequations}
\begin{align}
\label{red rep1}
\delta \mu _{1-g,p} = Q \, \mu _{-g,p} + \eta \, \mu _{-g,p+1} \, ,\hspace{5mm} (p = -1,0,1, \dots , g-1 ) \, , 
\end{align}
and the $g$-th gauge transformations preserving $(g-1)$-th gauge transformations of auxiliary gauge parameters 
\begin{align}
\label{red rep2}
\delta \mu _{1-g,p} = Q \, \mu _{-g,p} + \eta \, \mu _{-g,p+1} \, ,\hspace{5mm} (p < -1, \,  g-1 < p ) \, .  
\end{align}
\end{subequations} 
Note that all higher gauge parameters $\{ \mu _{-g,p} \} _{g>0, -1\leq p \leq g-1}$ and those of auxiliary gauge parameters $\{ \mu _{-g,p} \} _{g>0, p<-1,g\leq p}$ are independent and live in the large Hilbert space: 
\begin{align*}
\{ \mu _{-g , p } \} _{g>0 , p \in \mathbb{Z} } \subset \cH _{\xi \eta \phi } \, .
\end{align*}

\sub{Nonminimal set with constraints and free master action} 

The above analysis of the gauge reducibility implies that the set of gauge parameters $\{ \mu _{g,p} \} _{g<0, p\in \mathbb{Z} }$ appears in superstring field theory based on $\Psi , \delta \Psi \in \cH _{\beta \gamma }$\,. 
Hence, the set of fields--antifields is given by 
\begin{align}
\label{the minimal set}
\Psi _{1,-1} \in \cH _{\beta \gamma } \, , \hspace{3mm} 
\big{\{ } \Psi _{0,p} \big{ \} }_{p \in \mathbb{Z}} \subset 
 {\rm Ker} \big[ \eta \big] \cup {\rm Ker} \big[ Q \big] \, , \hspace{3mm} 
\big{\{ } \Psi _{-1-g,p} , \, (\Psi _{1-g,p} )^{\ast } \big{\} }_{g \geq 0, p\in \mathbb{Z}} \subset \cH  _{\xi \eta \phi } \, . 
\end{align}
We write $(\Psi _{g,p} )^{\ast }$ for the antifield of $\Psi _{g,p}$, whose ghost and picture numbers is determined via the BPZ inner product of  the theory as $\langle (\Psi _{g,p} )^{\ast } , \Psi _{g,p} \rangle \not= 0$. 
Note that the dynamical string field $\Psi _{1,-1} \in {\rm Ker} [\eta ]$ must satisfy the constraint 
\begin{subequations} 
\begin{align}
\eta \, \Psi _{1,-1} = 0 \, , 
\end{align}
and the first ghost fields $\Psi _{0,p} \in  {\rm Ker} \big[ \eta \big] \cup {\rm Ker} \big[ Q \big]$ must satisfy the constraints 
\begin{align}
Q \, \Psi _{0,-1} = \eta \, \Psi _{0,0} \, , \hspace{5mm} 
Q \, \Psi _{0,p} + \eta \, \Psi _{0,p+1} = 0 \,, \hspace{3mm} (p\not= -1) \, . 
\end{align}
\end{subequations} 
There is no constraint on the other fields and antifields: They live in the large Hilbert space. 

\vspace{1mm} 

The large form of the master action is given by the same form\footnote{Again, the property $\Psi _{0,-1} , \Psi _{0,0} \in  {\rm Ker} \big[ \eta \big] \cup {\rm Ker} \big[ Q \big]$ ($\subset {\rm Ker}[Q\eta ]$) is crucial for the master equation.} as (\ref{free master action}) 
\begin{align}
\label{redundant bv} 
S_{\sf bv} [\Psi , \Psi ^{\ast }] = \frac{1}{2} \la \xi \, \Psi _{1,-1} , \, Q \, \Psi _{1,-1} \ra 
+ \sum_{g=0}^{\infty } \sum_{p=-\infty }^{\infty } \la ( \Psi _{1-g,p})^{\ast } , \, Q \, \Psi _{-g,p} + \eta \, \Psi _{-g,1+p} \ra \,  
\end{align}
except for that the $p$-label runs over all integer numbers. 
It closely resembles to that of WZW-like theory \cite{Torii:2011zz, Kroyter:2012ni, Torii:2012nj, Iimori:2015aea}. 
One can find that it also reduces to the canonical form via canonical transformations as we proved in section 2. 
We can construct the canonical form of the master action for interacting theory in the same way as section 2. 

\subsection{The hidden gauge reducibility without constraints}  

We found that \textit{large} gauge symmetries are hidden in superstring field theory based on the small Hilbert space. 
However, because of the first class constraints (\ref{dep of mu1}-b), these large symmetries (\ref{re-expression}) and (\ref{red rep1}-b) are expressed in redundant way. 
We give a non-redundant expression of these large gauge symmetries and clarify the hidden gauge reducibility without using constraints. 


Let $\Lambda _{-1,0}$ be a string field of gauge parameters which lines in the large Hilbert space. 
The gauge transformation {free gauge transformation} is written as follows 
\begin{align}
\label{Psi}
\delta \Psi _{1,-1} = Q \, \eta \, \Lambda _{-1,0} \, . 
\end{align}
This is the origin of \textit{large} gauge symmetries arising from the \textit{small} theory. 
Clearly, this $\Lambda _{-1,0}$ equals to the half of gauge parameters appearing in the \textit{large} theory \cite{Kroyter:2012ni, Matsunaga:2018hlh}. 
This gauge transformation (\ref{Psi}) is invariant under the following gauge transformations  
\begin{align*}
\delta _{1} (\delta \Psi ) = 0 \, , \hspace{5mm} \delta _{1} \Lambda _{-1,0} = Q \, \Lambda _{-2,0} + \eta \, \Lambda _{-2,1} \, , 
\end{align*}
where $\Lambda _{-1-g,p}$ denotes a higher gauge parameter. 
Note that these $\Lambda _{-1-g,p}$ have the opposite Grassmann parity to $\mu _{-g,p}$ of (\ref{red rep1}-b). 
Likewise, we find the higher gauge transformations 
\begin{align}
\label{Lambda}
\delta _{g+1} ( \delta _{g} \Lambda _{-g,p} ) = 0 \, , 
\hspace{5mm} 
\delta _{g} \Lambda _{-g,p} = Q \, \Lambda _{-1-g,p} + \eta \, \Lambda _{-1-g,p+1} \, . 
\end{align}
The $p$-label of the higher gauge parameter $\Lambda _{-g,p}$ runs from $0$ to $g-1$, and thus, the $g=p$ line of $\Lambda _{-g,p}$ does not appear in these \textit{large} gauge symmetries of the \textit{small} theory. 

\sub{The minimal set and free master action}

Since these gauge parameters are Grassmann even unlike $\Psi \equiv \Psi _{1,-1}$, we write $\Phi _{-1-g,p}$ for the string field of ghosts corresponding to $\Lambda _{-1-g,p}$. 
We write $( \Psi _{1,-1})^{\ast }$ for the antifield of $\Psi $ and $( \Phi _{-1-g,p})^{\ast }$ for the antifield of $\Phi _{-1-g,p}$ respectively. 
These are defined by $\langle (\Phi _{\alpha } )^{\ast } , \Phi _{\alpha } \rangle = 1$ and their Grassmann parities satisfy $(-)^{\Psi } = (-)^{\Phi +1}$, $(-)^{\Psi ^{\ast } } = (-)^{\Psi +1}$, and $(-)^{\Phi ^{\ast }} = (-)^{\Phi +1}$. 
By counting the hidden gauge reducibility (\ref{Lambda}), we find the minimal set of fields--antifields 
\begin{align}
\label{minimal}
\Psi _{1,-1} \in \cH _{\beta \gamma } \, , \hspace{3mm} 
\big{\{ } \, \Phi _{-1-g,p} , \, (\Psi _{1,-1})^{\ast } \, , \, (\Phi _{-1-g,p} )^{\ast } \, \big{|} \, 0 \leq g \, , \, 0 \leq p \leq g \, \big{\} } \subset \cH  _{\xi \eta \phi } \, . 
\end{align}
Note that there is no constraint. 
The large gauge parameter (\ref{Psi}) enables us to obtain the minimal set of fields--antifields for the small theory with large gauge symmetries. 


We find that for the free theory, a proper BV master action is given by 
\begin{align}
\label{proper bv}
S_{\textsf{bv}} = \frac{1}{2} \la \, \xi \, \Psi \, , \, Q \, \Psi \, \ra  
+ \la \, (\Psi )^{\ast } , \, Q \, \eta \, \Phi _{-1,0} \, \ra 
+ \sum _{g=0}^{\infty }\sum_{p=0}^{g} 
\La (\Phi _{-1-g,p} )^{\ast } , \, Q \, \Phi _{-2-g,p} + \eta \, \Phi _{-2-g,p+1} \Ra \, . 
\end{align}
Clearly, this is nothing but a partially gauge-fixed version of the free master action for Berkovits theory \cite{Kroyter:2012ni, Matsunaga:2018hlh}. 
The origin of the \textit{large} gauge symmetries of the \textit{small} theory is the very trivial embedding of the small theory into the large Hilbert space (\ref{kinetic}), which seems to be trivial at the classical level but gives such results at the level of the master action. 


While one can apply some useful techniques developed in the previous section to the master action (\ref{redundant bv}) based on the nonminimal set with constraint, the master action (\ref{proper bv}) based on the minimal set (\ref{minimal}) necessitates the BV formalism in the large Hilbert space\cite{Matsunaga:2018hlh}. 
We thus focus on the former and give a recipe for interacting theories in the rest of this paper. 

\section{BV master action for interacting theory}

In the previous sections, we showed that there is a hidden gauge reducibility in superstring field theory based on the small dynamical string field $\Psi \in \cH _{\beta \gamma }$ whose gauge variation is also small $\delta \Psi \in \cH _{\beta \gamma }$. 
It requires additional propagating ghost--antighost fields in the gauge fixed or quantum gauge theory, and thus changes the set of BV fields--antifields. 
While the resultant master action can takes different and enlarged forms, there exist canonical transformations getting it back to the canonical form. 
In this section, on the basis of these results, we present master actions for several types of interacting superstring field theories. 

\subsection{Unrestricted fields--antifields in the canonical form} 

In the set of BV fields--antifields (\ref{primary constraints}), while the subspace of BV fields $\Psi _{g,p}$ is restricted by the constraints, there is no restriction on the subspace of BV antifields $\Phi _{2-g,-1-p}^{\ast } \equiv (\Psi _{g,p})^{\ast }$. 
It enables us to have various patterns of the master action and its canonical transformations. 
However, when we take various canonical transformations into account and focus on {\it the canonical form} of the master action $S_{\sf bv}$, these unrestricted antifields $\Phi ^{\ast } \in \cH _{\xi \eta \phi }$ all appear in the form of $\eta \, \Phi ^{\ast } \in \cH _{\beta \gamma }$ in $S_{\sf bv}$. 
Then, as we show, the master equation holds in a simple manner. 

We write $\Psi _{-}$ for the sum of all fields and $\Phi _{+}^{\ast }$ for the sum of all antifields. 
When the minimal set of fields--antifields is given by (\ref{primary constraints}), these $\Psi $ and $\Phi ^{\ast }$ take the following forms,  
\begin{align*} 
\Psi _{-} \equiv \Psi _{1,-1} + \sum_{g=0}^{\infty } \Big[ \Psi _{-g,-1} + \sum_{p=1}^{\infty } \Psi _{-g,-1-p} \Big] \,, 
\hspace{5mm}  
\Phi _{+}^{\ast } \equiv \Phi _{1,0}^{\ast } + \sum_{g=0}^{\infty } \Big[ \Phi ^{\ast }_{2+g,0} + \sum_{p=0}^{\infty }\Phi ^{\ast }_{2+g,p} \Big]  \, . 
\end{align*}
We set $\varphi \equiv \xi  \Psi _{-} + \Phi _{+}^{\ast }$ using these $\Psi _{-}$ and $\Phi _{+}^{\ast }$. 
When the original action $S[\Psi ]$ is given by (\ref{SFT action}), we can obtain the canonical form of the master action $S_{\sf bv}$ using this $\varphi $ by just replacing $\Psi $ of the original action $S[\Psi ]$ with $\eta \, \varphi $ as $S_{\sf bv} = S[\eta \varphi ]$, namely, 
\begin{align} 
\label{SFT master action} 
S_{\sf bv} = K \big( \eta \, \varphi , \, Q \, \eta \, \varphi \big)
+ \sum_{n\geq 3} V_{n} \big( \eta \, \varphi , \dots , \eta \, \varphi \big) 
+ \sum_{g} \sum_{n\geq 1} V_{g,n} \big( \underbrace{\eta \, \varphi , \dots , \eta \, \varphi }_{n} \big) \, .
\end{align}  
By construction, it quickly satisfies the classical master equation $\{ S_{\sf bv} , S_{\sf bv} \} = 0$ or the quantum master equation $\frac{1}{2} \{ S_{\sf bv} , S_{\sf bv} \} = \hbar \Delta S_{\sf bv}$ as the same manner as well-established theory based on the geometry-inspired restrictions. 
In particular, BV-BRST transformations of $\Psi _{-}$ and $\Phi _{+}^{\ast }$ are orthogonally split: $\delta _{\rm BV} \Psi _{-}$ is $\eta $-exact and $\delta _{\rm BV} \Phi _{+}^{\ast }$ is $\xi$-exact, which kills extra higher gauge symmetries. 
They are natural consequences of that we considered the canonical form. 

\vspace{2mm} 

Therefore, additional ghost--antighost string fields arising from the hidden gauge reducibility certainly propagate and contribute in the loop amplitudes of superstrings. 
Contribution of each ghost--antighost term will be changed via a gauge choice and canonical transformations. 
We thus expect that as usual gauge field theory, there exist appropriate gauge and suitable form of the master action for considering situations. 
For this purpose, the large class of canonical transformations and the large form of $S_{\sf bv}$ should be clarified. 
However, unfortunately, it remains unclear yet. 
We would like to emphasis that the above canonical form of $S_{\sf bv}$ will be canonical-transformed one from this unknown but large form of $S_{\sf bv}$. 
See also \cite{Matsunaga:2018hlh} for new results. 

\sub{Example: classical master action for open superstrings} 

In the rest of this subsection, we show it explicitly by taking open superstring field theory as an example. 
We consider the NS action for \cite{Witten:1986qs, Erler:2013xta, Konopka:2016grr} or the NS + R action\footnote{Then, as the inner product, we have to use that of \cite{Erler:2016ybs}. 
Note that $g$ and $p$ of $\Psi _{1-g,-1-p}$ denote $g$-th reducibility and $p$-decreasing from the natural picture number of considering string fields respectively.} for \cite{Erler:2016ybs}, 
\begin{align}
\label{action open}
S [ \Psi ] & = \frac{1}{2} \la \xi \, \Psi _{1,-1} , \, Q \, \Psi _{1,-1} \ra  + \sum_{n>1} \frac{1}{n+1} \la \xi \, \Psi _{1,-1} , \, M_{n} \big( \overbrace{ \Psi _{1,-1} , \dots , \Psi _{1,-1} }^{n} \big) \ra \, , 
\end{align}
where $M_{n}$ denotes the classical open superstring vertices $\{ V_{g,n} \} _{g=0,1}$ of (\ref{SFT master action}). 
The string field $\Psi $ and its gauge variation must satisfy (\ref{free constraint}). 
Then, $S_{\sf bv} = S [ \eta \varphi ]$ gives a solution of $\{ S_{\sf bv} , S_{\sf bv} \} = 0$. 
Using coalgebraic notation (See \cite{Erler:2014eba} for example.), we can express $S_{\sf bv}$ as 
\begin{align}
\label{master action open}
S_{\sf bv} [\Psi _{-} , \Phi _{+}^{\ast } ] &  
= \int _{0}^{1} dt \, \La \, \xi \Psi _{-} + \Phi _{+}^{\ast } \, , \, \pi _{1} \, \bM \frac{1}{1- t \, \eta \, ( \xi \Psi _{-} + \Phi _{+}^{\ast } ) } \, \Ra \, . 
\end{align}
Here, $t \in [0,1]$ is a real parameter. 
It derivatives are given by the following forms, 
\begin{align*} 
\frac{\partial _{r} S_{\sf bv} [\Psi _{-} , \Phi _{+}^{\ast } ]}{\partial \Psi _{1-g,-1-p}} 
= \pi_{1} \, \bxi \, \bM \frac{1}{1- \eta \, \varphi 
} \bigg{|}_{g,p}  \, , 
\hspace{3mm} 
\frac{\partial _{l} S_{\sf bv} [\Psi _{-} , \Phi _{+}^{\ast } ]}{\partial \Phi _{1+g,p}^{\ast } } 
= \pi_{1} \, \bM \frac{1}{1- \eta \, \varphi 
} \bigg{|}_{-g,-1-p} \, . 
\end{align*}
These are orthogonally split. 
In particular, the $A_{\infty }$ vertices $\bM $ acts on $\cH _{\beta \gamma }$ because all fields--antifields appear in the form of $\eta \, \varphi = \xi \, \Psi _{-} + \Phi _{+}^{\ast }$, which permits any $\xi $-assignment in the classical master equation. 
Because of $\eta = \eta \, \xi \, \eta$ and $\Eta \, \bM + \bM \, \Eta = 0$, we find 
\begin{align*}
\big{\{ } S_{\sf bv} , S_{\sf bv} \big{\} } = \LA  \pi_{1} \, \bM \frac{1}{1- \eta \, (\xi \, \Psi _{-} + \Phi _{+}^{\ast })} , \,  \pi_{1} \, \bxi \, \bM \frac{1}{1- \eta \, ( \xi \, \Psi _{-} +  \Phi _{+}^{\ast })} \RA = 0 \, . 
\end{align*}
The classical master action (\ref{master action open}) can be derived by induction based on the antifield expansion.

\sub{On-shell gauge reducibility and BV spectrum}

Let us check that (\ref{action open}) gives the same BV spectrum as (\ref{primary constraints}) before considering the construction of (\ref{master action open}). 
The action (\ref{action open}) has the gauge invariance under 
\begin{align}
\label{gauge transformation open}
\delta \Psi _{1,-1} & = \pi _{1} \bM \blambda _{0,-1} \frac{1}{1-\Psi _{1,-1} } 
\equiv Q \, \Lambda _{0,-1} + \sum_{n=1}^{\infty } \sum_{\rm cyclic} M_{n +1} \big( \overbrace{\Psi _{1,-1} , \dots , \Psi _{1,-1} }^{n} , \lambda _{0,-1} \big) \, , 
\end{align}
where $\blambda _{g,p}$ is the coderivation inserting the state $\lambda _{g,p}$. 
It is well-known that this gauge symmetry is {\it on-shell infinitely reducible}. 
\begin{subequations} 
Off-shell, there is no gauge reducibility and the gauge invariance necessitates $\lambda _{0,-1} \in \cH _{\beta \gamma }$, or equivalently 
\begin{align}
\label{off-shell lambda}
\eta \, \lambda _{0,-1} = 0 \, . 
\end{align}
We often write $\delta \Psi _{1,-1} = \cG \, \lambda _{0,-1}$ for (\ref{gauge transformation open}), and call $\cG$ (or $\bM$) as the gauge generator; this $\cG$ becomes nilpotent operator on-shell, which yields the on-shell gauge reducibility. 
We find that on-shell, the gauge parameter field $\Lambda _{0,-1}$ can protrude from the constraint (\ref{off-shell lambda}) as long as  
\begin{align*}
\lambda _{0,-1} = \eta \, \xi \, \lambda _{0,-1} + \sum_{n=1}^{\infty } \sum_{k=1}^{n} M_{n} \big( \overbrace{\Psi _{1,-1} , \dots }^{n-k} , \xi \, \lambda _{0,-2} , \overbrace{\Psi _{1,-1} , \dots }^{k-1} \big) \, , 
\end{align*}
where $\lambda _{0,-2}$ is an on-shell auxiliary gauge parameter fields. 
As well as (\ref{weaken constraint}), it leads a family of the {\it on-shell} auxiliary gauge parameter fields $\{ \lambda _{0,-1-p} \} _{p>0}$ satisfying the constraint 
\begin{align} 
\label{on-shell lambda} 
\eta \, \Lambda _{-p} + \pi _{1} \, \bM \frac{1}{1-\Psi _{1,-1}} \otimes \lambda _{0,-1-p} \otimes \frac{1}{1-\Psi _{1,-1}} = 0 \, .   
\end{align} 
\end{subequations} 
Hence, on-shell, we obtain $\{ \lambda _{0,-p} \} _{p>0} \subset \cH _{\beta \gamma } \oplus \cG \, \cH _{\xi \eta \phi }$ with $\delta \Psi _{1,-1} \equiv \cG \, \lambda _{0,-1}$ again. 

Next, let us consider the on-shell gauge transformations $\delta _{1}$ for the gauge transformation (\ref{gauge transformation open}), which must preserve (\ref{on-shell lambda}): 
For $p \geq 0$, they satisfy 
\begin{align*}
\pi _{1} \bM (\delta _{1} \blambda _{0,-1} ) \frac{1}{1-\Psi _{1,-1}} = 0 \, ,
\hspace{5mm} 
\eta \, ( \delta _{1} \lambda_{0,-p} ) + \pi _{1} \bM (\delta _{1} \blambda _{0,-1-p}) \frac{1}{1-\Psi _{1,-1}} = 0 \, . 
\end{align*}
Clearly, these yield the on-shell infinite gauge reducibility, and we find the $g$-th gauge transformations $\delta _{g}$ which preserve the $(g-1)$-th gauge transformations $\delta _{g} ( \delta _{g-1} \lambda _{1-g,-1-p} ) = 0$, 
\begin{align}
\label{lgs int}
\delta _{g} \lambda _{1-g,-1} & = \pi _{1} \bM \blambda _{-g,-1} \frac{1}{1-\Psi _{1,-1} } , 
\hspace{3mm} 
\delta _{1} \lambda _{1-g,-1-p}  = \eta \, \lambda _{-1,-p} + \pi _{1} \bM \blambda _{-1,-1-p} \frac{1}{1-\Psi _{1,-1} } .  
\end{align}
Therefore, as we found in the previous section, in addition to $\Psi _{1,-1}$ and $\lambda _{1,-1}$, an infinite tower of higher on-shell gauge parameters (\ref{lgs int}) appear in the interacting theory. 
Hence, the set of fields--antifields is given by the same BV spectrum as (\ref{primary constraints}). 

\sub{General form of the antifield number expansion}

We construct a BV master action which consists of the BV spectrum (\ref{primary constraints}). 
Again, we consider the antifield number expansions of the master action (\ref{afn expansion}) and the master equation, 
\begin{align*}
\big{\{ } S_{\sf bv} , S_{\sf bv} \big{\} } & = \sum_{a=0}^{\infty } \big{\{ } S_{\sf bv} , S_{\sf bv} \big{\} } \big{|}^{(a)} \, . 
\end{align*}
In this case, the initial condition of the BV master action, $S^{(0)} [ \psi ] \equiv S [\Psi ]$, is given by (\ref{action open}). 
For this purpose, we derive the explicit form of the antifield number $a$ part of the master equation $\{ S_{\sf bv} , S_{\sf bv} \} = 0$\,. 
Then, we find that only the perturbative solutions $\{ S^{(n)} \} _{n=0}^{a+1}$ up to the antifield number $(a+1)$ appear in the antifield number $a$ part of the master equation, 
\begin{align*} 
\big{\{ } S_{\sf bv} , S_{\sf bv} \big{\} } \big{|}^{(a)} & = \big{\{ } S^{(0)} + \dots + S^{(a+1)} , \, S^{(0)} + \dots + S^{(a+1)} \big{\} } \, , 
\end{align*}
which is one of powerful properties of the antifield expansion in string field theory. 
When we consider the BV master action $S_{\sf BV}[\Psi , \Phi ^{\ast } ]$ based on (\ref{primary constraints}), by the assignment of the antifield number and space-time ghost number, each $S^{(a)}$ must satisfy the following relations 
\begin{align*} 
\frac{\partial S^{(1+a)} }{\partial \Phi ^{\ast }_{1+g,p} } = \frac{\partial S^{(a)} }{\partial \Psi _{1-g,-1-p} } = 0 \, , \hspace{5mm} ( g > a ) \, .
\end{align*}
They completely determine the explicit form of the antifield number expansion of the master equation. 
Note that, by construction, ${\rm afn} [ S^{(a)} ] = a$ holds. 
Because of ${\rm afn} [ \Phi ^{\ast }_{1+g,p} ] = 1+g $ and ${\rm afn} [ \Psi _{1-g,-1-p} ] = 0$ by definition, the antifield number of devatives are assigned as 
\begin{align*}
{\rm afn} \bigg[ \frac{\partial S^{(1+a)} }{\partial \Phi ^{\ast }_{1+g,p} } \bigg] & = a - g \, , 
\hspace{5mm} 
{\rm afn} \bigg[ \frac{\partial S^{(a)} }{\partial \Psi _{1-g,-1-p} } \bigg] = a \, . 
\end{align*}
We thus obtain the following relation in the antifield number expansion,   
\begin{align*}
\sum_{a = 0}^{\infty } \big{ \{ } S_{\sf bv} , \, S_{\sf bv} \big{\} } \big{|}^{(a)} 
= \sum_{a=0}^{\infty } \sum_{t,s} \sum_{g,p} \LA \frac{\partial _{r} S^{(t)}}{\partial \Psi _{1-g,-1-p}} , \, \frac{\partial _{l} S^{(s)} }{\partial \Phi ^{\ast }_{1+g,p}} \RA \bigg{|}_{t+s-g-1 =a} \, . 
\end{align*}
The antifield number $a$ terms are given by 
\begin{align}
\label{explicit} 
\big{ \{ } S_{\sf bv} , \, S_{\sf bv} \big{\} } \big{|}^{(a)} %
& = \sum_{s=0}^{a} \sum_{g=0}^{s} \sum_{p=0}^{\infty } \LA \frac{\partial _{r} S^{(a-[s-g])}}{\partial \Psi _{1-g,-1-p}} , \, \frac{\partial _{l} S^{(1+s)} }{\partial \Phi ^{\ast }_{1+g,p}} \RA \, , 
\end{align}
and we find that $\{ S^{(n)} \} _{n>a+1}$ do not appear in $\{ S_{\sf bv} , S_{\sf bv} \} |^{(a)}=0$\,.

\sub{Regarding ambiguity of $\xi $-assignments and lowest solution} 

We start with the initial condition $S^{(0)} \equiv S [ \Psi ]$ given by (\ref{action open}). 
First, we would like to specify the lowest solution $S^{(1)} = S^{(1)} [ \Psi , \Phi ^{\ast } ]$ which has to satisfy 
\begin{align*}
\frac{1}{2} \Big{\{ } \, S_{\sf bv} [\Psi , \Phi ^{\ast }] \, , \, S_{\sf bv}[\Psi , \Phi ^{\ast } ] \, \Big{\} } \Big{|}^{(0)} 
& = \frac{1}{2} \Big{\{ } \, S^{(0)} + S^{(1)} , \, S^{(0)} + S^{(1)} \, \Big{\} } 
= \LA \, \frac{\partial S^{(0)} }{\partial \Psi _{1,-1}} \, , \, \frac{\partial S^{(1)} }{\partial \Phi ^{\ast }_{1,0}} \, \RA 
= 0 \, . 
\end{align*}
To find an appropriate $S^{(1)}$, we have to determine these derivatives. 
Note that because of the constraint $\eta \Psi _{1,-1} = 0$, the action $S^{(0)} [\Psi ] = S [\Psi ]$ permits any $\xi $-assignment: 
Using real parameters $t_{0} , \dots , t_{n} \in \mathbb{R}$ satisfying $t_{0} + \dots + t_{n} = 1$\,, we can express it as 
\begin{align*}
S^{(0)} [\Psi ] 
& = \sum_{n=1}^{\infty} \frac{1}{n+1} 
\Big[ t_{0} \, A_{n,0} [\Psi ] + t_{1} \, A_{n,1} [\Psi ] + \dots + t_{n} \, A_{n,n} [\Psi ] \Big] \, , 
\end{align*}
where $A_{n,0}$ and $A_{n,k}$ for $k=1,\dots ,n$ are defined by their $\xi $-assignments in $M_{n}$\,, 
\begin{align*}
A_{n,0} [\Psi ] & \equiv \la \Psi _{1,-1} , \, \xi \, M_{n} \big( \overbrace{\Psi _{1,-1} , \dots , \Psi _{1,-1} }^{n} \big) \ra  \, , 
\\ 
A_{n,k} [\Psi ] & \equiv - \la \Psi _{1,-1} , \, M_{n} \big( \overbrace{\Psi _{1,-1} , \dots , \Psi _{1,-1} }^{k-1} , \xi \, \Psi _{1,-1} ,  \overbrace{\Psi _{1,-1} , \dots , \Psi _{1,-1} }^{n-k} \big) \ra  \, . 
\end{align*}
Hence, one can choose the right derivative of $S^{(0)}$ as 
\begin{align*}
- \LA \frac{\partial _{r} S^{(0)}[\Psi ]}{\partial \Psi _{1,-1}} , \, \Psi _{1,-1} \RA 
& = t_{0} \, A_{0} [\Psi ] + t_{1} A_{1} [\Psi ] + \dots + t_{n} \, A_{n} [\Psi ] \, .
\end{align*}
This ambiguity may enable us to construct various forms of $S_{\sf bv}$ or give some hint to obtain a large class of consistent master actions. 
However, we consider the simplest case in this paper. 
Since $\eta A_{0} [\Psi ] = \dots = \eta A_{n} [ \Psi ]$ holds by acting $\eta $ on these, we find an unambiguous expression 
\begin{align*} 
\Eta \, \frac{\partial _{r} S^{(0)}[\Psi ]}{\partial \Psi _{1,-1}} & = \pi _{1} \, \bM \frac{1}{1-\Psi _{1,-1} } 
= \sum_{n=1}^{\infty } M_{n} \big( \overbrace{\Psi _{1,-1} , \dots , \Psi _{1,-1} }^{n} \big) \, . 
\end{align*}
Thus if we set the following left derivative of $S^{(1)}$,   
\begin{align*}
\frac{\partial _{l} S^{(1)}}{\partial \Phi ^{\ast }_{1,0}} 
& = - \pi _{1} \, \bM \, ( \Eta \, \bxi \, \bPsi _{0,-1}) \frac{1}{1- \Psi _{1,-1} }  \, , 
\end{align*} 
it always satisfies the antifield number $(0)$ part of the master equation with any $\frac{\partial S^{(0)}}{\partial \Psi _{1,-1}}$\,. 
Here, $\Eta \, \bxi \, \bPsi _{0,-1}$ denotes the coderivation inserting $\eta \, \xi \, \Psi _{0,-1}$ into the tensor algebra of $\Psi \equiv \Psi _{1,-1}$. 
Then, we obtain the lowest solution $S^{(1)}$ for $\{ S^{(0)} + S^{(1)} , \, S^{(0)} + S^{(1)} \} = 0$ as 
\begin{align*}
S^{(1)} & = \La \, \Phi ^{\ast }_{1,0} \, , \, \pi _{1} \, \bM \, ( \Eta \, \bxi \, \bPsi _{0,-1}) \frac{1}{1-\Psi _{1,-1}} \, \Ra \, .
\end{align*}
In general, the antifield number expansion does not uniquely determine the form of the master action because one can consider various canonical transformations at each order: 
It would be an interesting to consider all possible transformations of $S^{(0)} + \dots + S^{(a)}$. 
However, we consider this simplest form of $S^{(1)}$, the same form as well-established theory, and we would like to focus on the canonical form in this paper, which makes analysis very simple. 

\sub{Inductive construction of the simplest solution} 

In the rest, we consider the construction of the master action which takes the canonical form. 
Thus, we omit the projector $\Eta \, \bxi $ in front of the coderivation $\Eta \, \bxi \, \bPsi _{g,p}$ and write $\bPsi _{g,p}$ for brevity. 
At this step, we have the following right derivatives 
\begin{align*} 
\frac{\partial _{r} S^{(1)} }{\partial \Psi _{0,-1}} 
& = \pi _{1} \, \bM \, \bPhi ^{\ast }_{1,0} \frac{1}{1-\Psi _{1,-1}} + \eta \mathchar`- {\rm exact} \, , 
\\  
\frac{\partial _{r} S^{(1)} }{\partial \Psi _{1,-1}} 
& = \pi _{1} \, \bM \, \bPhi ^{\ast }_{1,0} \, \bPsi _{0,-1} \frac{1}{1-\Psi _{1,-1}} + \eta \mathchar`- {\rm exact} \, .
\end{align*} 
We set $\frac{\partial _{r} S^{(1)}}{\partial \Psi _{0,-p}}= 0$ for $p>1$. 
Thus, to solve the antifield number $1$ part of the equation, 
\begin{align*}  
\big{ \{ } S_{\sf BV} , \, S_{\sf BV} \big{\} } \big{|}^{(1)} 
& = \sum_{p=0}^{\infty } \LA \frac{\partial _{r} S^{(1)} }{\partial \Psi _{0,-1-p}} , \, \frac{\partial _{l} S^{(2)} }{\partial \Phi ^{\ast }_{2,p}} \RA  
+ \LA \frac{\partial _{r} S^{(0)}}{\partial \Psi _{1,-1}} , \, \frac{\partial _{l} S^{(2)} }{\partial \Phi ^{\ast }_{1,0}} \RA  
+ \LA \frac{\partial _{r} S^{(1)}}{\partial \Psi _{1,-1}} , \, \frac{\partial _{l} S^{(1)} }{\partial \Phi ^{\ast }_{1,0}} \RA \, , 
\end{align*}
we need the following terms 
\begin{align*}
\frac{1}{2} \La \, \bM \, \bPhi _{1,0}^{\ast } \, , \, \bM \, \bPsi _{0,-1} \, \bPsi _{0,-1} \, \Ra 
+ \frac{1}{2} \La \, \bM \, \bPhi _{1,0}^{\ast } \, \bPsi _{0,-1} \, \bPsi _{0,-1} \, , \, \bM \, \Ra \, . 
\end{align*}
These terms should be provided by the inner product of the above field derivatives and the next antifield derivatives. 
Therefore, the left derivatives of $S^{(2)}$ should be 
\begin{align*}
\frac{\partial _{l} S^{(2)}_{0}}{\partial \Phi ^{\ast }_{2,p}} 
& = \pi _{1} \, \bM \Big[ \bPsi _{-1,-1-p} + \sum_{q=0}^{p} \bPsi _{0,-1-q} \bPsi _{0,-1-(p-q)} \Big] \frac{1}{1-\Psi _{1,-1}} \, , 
\\ 
\frac{\partial _{l} S^{(2)}_{0}}{\partial \Phi ^{\ast }_{1,0}} 
& = \pi _{1} \, \bM ( \Eta \bPhi ^{\ast }_{1,0} ) \Big[ \bPsi _{-1,-1} + \frac{1}{2} ( \bPsi _{0,-1})^{2} \Big] \frac{1}{1-\Psi _{1,-1}}   \, . 
\end{align*}
We thus find that the antifield number $2$ part is given by 
\begin{align*}
S^{(2)} 
& = \sum_{p=0}^{\infty } \LA \Phi ^{\ast }_{2,p} , \, \pi _{1} \, \bM  \bigg[ \bPsi _{-1,-1-p} + \frac{1}{2} ( \bPsi _{0,-1} )^{2} + \sum_{q=0}^{p} \bPsi _{0,-1-q} \bPsi _{0,-1-(p-q)} \bigg] \frac{1}{1- \Psi _{1,-1} } \RA \, 
\no & \hspace{15mm}  
+  \frac{1}{2} \LA \eta \, \Phi _{1,0}^{\ast } , \, \pi _{1} \, \bM  \bPhi _{1,0}^{\ast } \Big[ \bPsi _{-1,-1} + \frac{1}{2} ( \bPsi _{0,-1} )^{2} \Big] \frac{1}{1- \Psi _{1,-1} } \RA \, . 
\end{align*}
The important result at this step is that when we set 
\begin{align*} 
\Psi _{n} \equiv \Psi _{1,-1} + \sum_{g=0}^{n-1} \Big[ \Psi _{-g,-1} + \sum_{p=1}^{\infty } \Psi _{-g,-1-p} \Big] \,, 
\hspace{5mm}  
\Phi _{n}^{\ast } \equiv \Phi _{1,0}^{\ast } + \sum_{g=0}^{n-1} \Big[ \Phi ^{\ast }_{2+g,0} + \sum_{p=0}^{\infty }\Phi ^{\ast }_{2+g,p} \Big]  \, , 
\end{align*}
we can obtain the following expression for $n=2$\,, 
\begin{align}
\label{recursive action}
S^{(0)} + \dots + S^{(n)} = \int _{0}^{1} dt \, \La \, \xi ( \Psi _{n} + \eta \, \Phi _{n-1}^{\ast } ) \, , \, \pi _{1} \, \bM \frac{1}{1- t \, \eta \, ( \xi \Psi _{n} + \Phi _{n-1}^{\ast } ) } \, \Ra 
\, . 
\end{align}

Using this $S^{(2)}$, one can determine the derivatives of $S^{(3)}$. 
For example, we have the $p=0$ right derivatives 
\begin{align*}
\frac{\partial _{r} S^{(2)}_{0}}{\partial \Psi _{-1,-1}} 
& = \pi _{1} \, \bM \Big[ \bPhi _{2,0}^{\ast } + \frac{1}{2} ( \Eta \, \bPhi _{1,0}^{\ast } ) \, \bPhi _{1,0}^{\ast } \Big] \frac{1}{1-\Psi _{1,-1}} 
+ \eta \mathchar`- {\rm exact} \, , 
\\ 
\frac{\partial _{r} S^{(2)}_{0}}{\partial \Psi _{-0,-1}} 
& = \pi _{1} \, \bM \Big[ \bPhi _{2,0}^{\ast } + \frac{1}{2} (\Eta \, \bPhi _{1,0}^{\ast } ) \, \bPhi _{1,0}^{\ast } \Big] \bPsi _{0,-1} \frac{1}{1-\Psi _{1,-1}} 
+ \eta \mathchar`- {\rm exact} \, , 
\\ 
\frac{\partial _{r} S^{(2)}_{0}}{\partial \Psi _{1,-1}} 
& = \pi _{1} \, \bM \Big[ \bPhi _{2,0}^{\ast } + \frac{1}{2} (\Eta \, \bPhi _{1,0}^{\ast } ) \, \bPhi _{1,0}^{\ast } \Big] \big( \bPsi _{-1,-1 } + \frac{1}{2} ( \bPsi _{0,-1} )^{2} \big) \frac{1}{1-\Psi _{1,-1}} 
+ \eta \mathchar`- {\rm exact} \, . 
\end{align*}
Note that although it seems that the second terms are enough to satisfy the lower master equation, the first terms of these derivative are necessitated to fix the on-shell gauge reducibility of $S^{(0)}$\,, which satisfy the lower master equation individualy. 
The antifield number $2$ part is 
\begin{align*} 
\big{ \{ } S_{\sf bv} , \, S_{\sf bv} \big{\} } \big{|}^{(2)} 
& = \sum_{p=0}^{\infty } \bigg[ \LA \frac{\partial _{r} S^{(2)}}{\partial \Psi _{-1,-1-p}} , \, \frac{\partial _{l} S^{(3)} }{\partial \Phi ^{\ast }_{3,p}} \RA \, 
+ \LA \frac{\partial _{r} S^{(1)}}{\partial \Psi _{0,-1-p}} , \, \frac{\partial _{l} S^{(3)} }{\partial \Phi ^{\ast }_{2,p}} \RA \bigg] 
+ \LA \frac{\partial _{r} S^{(0)}}{\partial \Psi _{1,-1}} , \, \frac{\partial _{l} S^{(3)} }{\partial \Phi ^{\ast }_{1,0}} \RA 
\no & \hspace{5mm} 
+ \sum_{p=0}^{\infty } \LA \frac{\partial _{r} S^{(2)}}{\partial \Psi _{0,-1-p}} , \, \frac{\partial _{l} S^{(2)} }{\partial \Phi ^{\ast }_{2,p}} \RA \,
+ \LA \frac{\partial _{r} S^{(1)}}{\partial \Psi _{1,-1}} , \, \frac{\partial _{l} S^{(2)}}{\partial \Phi ^{\ast }_{1,0}} \RA  
+ \LA \frac{\partial _{r} S^{(2)}}{\partial \Psi _{1,-1}} , \, \frac{\partial _{l} S^{(1)}}{\partial \Phi ^{\ast }_{1,0}} \RA \, . 
\end{align*}
Therefore, for example, we find that the $p=0$ slice of the second line requires  
\begin{align*}
&
 \LA \pi _{1} \, \bM \Big[ \bPhi _{2,0}^{\ast } + \frac{1}{2} (\Eta \bPhi _{1,0}^{\ast } ) \bPhi _{1,0}^{\ast } \Big] \bPsi _{0,-1} \frac{1}{1-\Psi _{1,-1}} 
, \, 
\pi _{1} \,\bM \Big[ \bPsi _{-1,-1 } + \frac{1}{2} ( \bPsi _{0,-1} )^{2} \Big] \frac{1}{1-\Psi _{1,-1}} \RA \, 
\no & \hspace{2mm}
+ \LA \pi _{1} \, \bM \bPhi ^{\ast }_{1,0} \bPsi _{0,-1} \frac{1}{1-\Psi _{1,-1}} 
, \, 
\pi _{1} \, \bM ( \Eta \bPhi ^{\ast }_{1,0} ) \Big[ \bPsi _{-1,-1} + \frac{1}{2} ( \bPsi _{0,-1})^{2} \Big] \frac{1}{1-\Psi _{1,-1}}  \RA \, 
\no & \hspace{5mm} 
+ \LA \pi _{1} \, \bM \Big[ \bPhi _{2,0}^{\ast } + \frac{1}{2} (\Eta \bPhi _{1,0}^{\ast } ) \bPhi _{1,0}^{\ast } \Big] \big( \bPsi _{-1,-1 } + \frac{1}{2} ( \bPsi _{0,-1} )^{2} \big) \frac{1}{1-\Psi _{1,-1}} 
, \, 
\pi _{1} \, \bM  \bPsi _{0,-1} \frac{1}{1-\Psi _{1,-1}} \RA  \, . 
\end{align*}
It implies that the $p=0$ left derivatives of $S^{(3)}$ should be 
\begin{align*}
\frac{\partial _{l} S^{(3)}}{\partial \Phi ^{\ast }_{3,0}} 
& = \pi _{1} \, \bM \Big[ \bPsi _{-2,-1} + \bPsi _{-1,-1} \bPsi _{0,-1} + \frac{1}{3!} ( \bPsi _{0,-1} )^{3} \Big] \frac{1}{1-\Psi _{1,-1}} 
\\  
\frac{\partial _{l} S^{(3)}}{\partial \Phi ^{\ast }_{2,0}} 
& = \pi _{1} \, \bM ( \Eta \bPhi _{1,0}^{\ast } ) \Big[ \bPsi _{-2,-1} + \bPsi _{-1,-1} \bPsi _{0,-1} + \frac{1}{3!} ( \bPsi _{0,-1} )^{3} \Big] \frac{1}{1-\Psi _{1,-1}} 
\\  
\frac{\partial _{l} S^{(3)}}{\partial \Phi ^{\ast }_{1,0}} 
& = \pi _{1} \, \bM \big( \Eta \bPhi _{2,0}^{\ast } + \frac{1}{2} (\Eta \bPhi _{1,0} ^{\ast })^{2} \big) \Big[ \bPsi _{-2,-1} + \bPsi _{-1,-1} \bPsi _{0,-1} + \frac{1}{3!} ( \bPsi _{0,-1} )^{3} \Big] \frac{1}{1-\Psi _{1,-1}}   \, . 
\end{align*}
In the same manner, one can obtain the $p>0$ derivatives by arranging the combination of coderivations $\bPsi _{-g,-p}$ and $\Phi _{g,p}^{\ast }$ in the above $p=0$ ones. 
These derivatives determine $S^{(3)}$, and this $S^{(3)}$ gives (\ref{recursive action}) for $n=3$\,. 
Inductively, one can prove that $S^{(n+1)}$ satisfies (\ref{recursive action}) for $n+1$ when $S^{(n)}$ satisfying it. 
Hence, we obtain the canonical form of the classical master action $S_{\sf bv}$ as the $n \rightarrow \infty $ limit of (\ref{recursive action}).

\subsection{BV master actions for superstring field theories} 

We found that by adding higher $p$-labeled fields--antifields into the known minimal set, by considering the sum of all fields $\Psi _{-}$ and antifields $\Phi _{+}^{\ast }$, and by setting $\varphi \equiv \xi \Psi _{-} + \Phi _{+}^{\ast }$, {\it the canonical form} of the master action $S_{\sf bv}$ including these additional propagating fields--antifields is obtained by just replacing $\Psi $ of the known action $S[\Psi ]$ with $\eta \varphi $, namely, $S_{\sf bv} = S [\eta \varphi ]$. 

\sub{Open superstring field theories} 

We know classical master actions $S[\widetilde{\Psi }]$ for geometrical open superstring field theories \cite{Konopka:2016grr}, and their homotopy algebraic versions \cite{Erler:2013xta, Erler:2016ybs}. 
From the analysis of the gauge reducibility of \cite{Erler:2013xta, Erler:2016ybs}, we find that the minimal set of fields--antifields can be enlarged as 
\begin{align*}
\big{\{ } \Psi _{1-g,-1-p}^{\rm NS} , \Phi _{1+g,p}^{\ast \, {\rm NS}} \equiv ( \Psi _{1-g,-1-p}^{\rm NS} )^{\ast } ; \Psi _{1-g, -\frac{1}{2}-p}^{\rm R} , \Phi _{1+g,\frac{1}{2} + p}^{\ast \, {\rm R}} \equiv ( \Psi _{1-g, -\frac{1}{2}-p}^{\rm R} )^{\ast } \big{\} }_{g,p\geq 0} \, . 
\end{align*}
Note that except for $\Psi _{1,-1}^{\rm NS} , \Psi _{1,-\frac{1}{2}}^{\rm R} \in \cH _{\beta \gamma }$ and $\{ \Psi _{0,-p}^{\rm NS/R} \} \subset \cH _{\beta \gamma } \oplus \cG \, \cH _{\xi \eta \phi }$, the other fields--antifields are unrestricted $\{ \Psi , \Phi ^{\ast } \} \subset \cH _{\xi \eta \phi }$. 
Therefore, by setting $\varphi \equiv \xi \Psi _{-} + \Phi _{+}^{\ast }$ where  
\begin{align*}
\Psi _{-} & \equiv \Psi _{1,-1}^{\rm NS} + \Psi _{1,-\frac{1}{2}}^{\rm R} + \sum _{g=0}^{\infty } \sum_{p=0}^{\infty } \Big[ \Psi _{-g,-p}^{\rm NS} + \Psi _{-g,\frac{1}{2} - p}^{\rm R} \Big] \, , 
\no 
\Phi _{+}^{\ast } & \equiv \Phi _{1,0}^{\ast \, {\rm NS} } + \Phi _{1,\frac{1}{2} }^{\ast \,{\rm R}} 
+ \sum _{g=0}^{\infty } \sum_{p=0}^{\infty } \Big[ \Phi _{1+g,p}^{\ast \, {\rm NS}} + \Phi _{1+g,\frac{1}{2} + p}^{\ast \, {\rm R}} \Big] \, , 
\end{align*}
we can quickly obtain the canonical form of the master action $S_{\sf bv}$ by just replacing $\widetilde{\Psi }$ of the action $S[\widetilde{\Psi }]$ given in \cite{Erler:2016ybs} with $\eta \varphi $, namely $S_{\sf bv} = S[ \eta \varphi ]$. 
(Note that there is the $Y$-insertion in the inner product of $R$ states.) 
For \cite{Konopka:2016grr}, we consider the set of fields--antifields 
\begin{align*}
\big{\{ } \Psi _{1-g,-1-p}^{\rm NS} , \Phi _{1+g,p}^{\ast \, {\rm NS}} ; \Psi _{1-g, -\frac{1}{2}-p}^{\rm R} , \Phi _{1+g,-\frac{1}{2} + p}^{\ast \, {\rm R}} \equiv ( \Psi _{1-g, -\frac{1}{2}-p}^{\rm R} )^{\ast } , \widetilde{\Psi }_{1-g,-\frac{3}{2}-p}^{\rm R} , ( \widetilde{\Psi }_{1-g,-\frac{3}{2}-p}^{\rm R} )^{\ast } \big{\} }_{g,p\geq 0} \, . 
\end{align*}
We set $\varphi \equiv \xi \Psi _{-} + \Phi _{+}^{\ast} $ and $\widetilde{\psi } \equiv \xi \widetilde{\Psi }_{-} + \widetilde{\Psi }_{+}^{\ast }$ where 
\begin{align*}
\Psi _{-} & \equiv \Psi _{1,-1}^{\rm NS} + \Psi _{1,-\frac{1}{2}}^{\rm R} + \sum _{g=0}^{\infty } \sum_{p=0}^{\infty } \Big[ \Psi _{-g,-p}^{\rm NS} + \Psi _{-g,\frac{1}{2} - p}^{\rm R} \Big] \, , 
\hspace{4mm}  
\widetilde{\Psi }_{-}  \equiv  \widetilde{\Psi }_{1,-\frac{3}{2} }^{\rm R} 
+ \sum _{g=0}^{\infty } \sum_{p=0}^{\infty } \widetilde{\Psi }_{1-g, -\frac{3}{2} - p}^{\rm R} \, , 
\\ 
\Phi _{+}^{\ast } & \equiv \Phi _{1,0}^{\ast \, {\rm NS} } + \Phi _{1,-\frac{1}{2} }^{\ast \,{\rm R}} 
+ \sum _{g=0}^{\infty } \sum_{p=0}^{\infty } \Big[ \Phi _{1+g,p}^{\ast \, {\rm NS}} + \Phi _{1+g, -\frac{1}{2} + p}^{\ast \, {\rm R}} \Big] \, , 
\hspace{2mm}    
\widetilde{\Psi }_{+}^{\ast }  \equiv  (\widetilde{\Psi }_{1,-\frac{3}{2} }^{\rm R} )^{\ast }  
+ \sum _{g=0}^{\infty } \sum_{p=0}^{\infty } ( \widetilde{\Psi }_{1-g, -\frac{3}{2} - p}^{\rm R} )^{\ast} \, . 
\end{align*}
Then, the enlarged BV master action $S_{\sf bv}$ is obtained by just replacing $(\Psi , \psi )$ and $\phi $ of $S[ (\Psi , \psi ) ; \phi ]$ given in \cite{Konopka:2016grr} with $\eta \varphi $ and $\eta \widetilde{\psi }$ respectively: $S_{\sf bv } = S [ \eta \varphi ; \eta \widetilde{\psi } ]$. 

\sub{Type II and Heterotic theories} 

We have quantum master actions $S_{q} [ \Psi , \widetilde{\Psi } ]$ for type II and heterotic superstring field theories \cite{Jurco:2013qra, Sen:2015uaa}, and their classical and homotopy algebraic versions \cite{Erler:2014eba}. 
We consider the quantum master action $S_{q}$ of \cite{Sen:2015uaa}. 
Then, the analysis of its gauge gauge reducibility implies that one can introduce the additional fields--antifields 
\begin{align*}
\big{\{ } \Psi _{-g,-1-p}^{\rm NS} , \Phi _{2+g,p}^{\ast \, {\rm NS}} ; \Psi _{-g, -\frac{1}{2}-p}^{\rm R} , \Phi _{2+g,-\frac{1}{2} + p}^{\ast \, {\rm R}} , \widetilde{\Psi }_{-g,-\frac{3}{2}-p}^{\rm R} , ( \widetilde{\Psi }_{-g,-\frac{3}{2}-p}^{\rm R} )^{\ast } \big{\} }_{g,p\geq 0} \, . 
\end{align*} 
Let $\Psi _{-}$ and $\Phi _{+}^{\ast }$ be the sums of all fields and antifields without ``tilde'', respectively, and let $\widetilde{\Psi }_{-}$ and $\widetilde{\Psi }_{+}^{\ast }$ be the sums of all fields and antifields with``tilde'', respectively. 
Using these, we set $\varphi \equiv \xi \Psi _{-} + \Phi _{+}^{\ast }$ and $\widetilde{\psi } \equiv \xi \widetilde{\Psi }_{-} + \widetilde{\Psi }_{+}^{\ast }$. 
Then, the quantum BV master action $S_{\sf bv}$ is obtained by just replacing $\Psi $ and $\widetilde{\Psi }$ of $S_{q} [\Psi , \widetilde{\Psi } ]$ given in \cite{Sen:2015uaa} with $\eta \varphi $ and $\eta \widetilde{\psi }$ respectively; $S_{\sf bv} = S_{q} [ \eta \varphi , \eta \widetilde{\psi }]$.

\section{Concluding remarks}

In this paper, we showed that there exists a hidden gauge reducibility in superstring field theory based on $\Psi , \delta \Psi \in \cH _{\beta \gamma }$. 
It necessitates additional propagating ghost--antighost fields in the gauge fixed or quantum gauge theory, and thus changes the set of fields--antifields. 
In terms of a gauge theory, it corresponds to hundle higher gauge symmetry which is fixed or ignored so far. 
We proved that the resultant master action can takes a different and enlarged form, and that canonical transformations fills their gap. 

We also checked that these additional fields--antifields can be put into the master actions for the interacting theories. 
Hence, these {\it additional propagating degrees of freedom} indeed appear in loop amplitudes of superstring field theory. 
It is known that we sometime encounter singular situations, such as spurious poles, in usual loop calculations \cite{deLacroix:2017lif, 
Erler:2017pgf}. 
Our analysis of the gauge structure implies that one can include additional contributions for loops {\it via the gauge choice}. 
Thus, it will be an interesting question whether one can control such singularities appearing in the loop superstring amplitudes via the gauge invariance of the field theory. 

Since the set of fields--antifields is enlarged, one cannot obtain unique $S_{\sf bv}$ by just relaxing the ghost number constraint, unlike usual cases. 
It implies the existence of a larger class of consistent Batalin-Vilkovisky master actions for superstring field theory. 
We presented it explicitly for free theory, and gave not all but several exact results for interacting theory. 
Interestingly, it remind us the WZW-like formulation of superstring field theory \cite{Berkovits:1995ab, Berkovits:2004xh, Matsunaga:2014wpa, Erler:2015uoa, Kunitomo:2015usa, Matsunaga:2015kra, Goto:2015pqv, GK, Matsunaga:2016zsu, Erler}. 

\vspace{1mm}

What is the origin of the hidden gauge reducibility and these additional loop-propagating degrees? 
It will be the ambiguity appearing in the expressions of the gauge transformations (for gauge transformations) in superstring field theory based on $\Psi , \delta \Psi \in \cH _{\beta \gamma }$. 
In section 3.1, we considered another gauge parameter $\omega _{0,0} \in {\rm Ker} [\eta ] \cup {\rm Ker} [Q]$ such that $Q \, \lambda _{0,-1} = \eta \, \omega _{0,0}$\,, which provides another expression $\delta \Psi _{1,-1} = \eta \, \omega _{0,0}$ of the gauge transformation $\delta \Psi _{1,-1} = Q \, \lambda _{0,-1}$ with $\lambda _{0,-1} \in {\rm Ker} [\eta ] \cup {\rm Ker} [Q]$\,. 
Although it is just a redefinition of gauge parameters because of $\delta \Psi _{1,-1} \in \cH _{\beta \gamma }$, as we showed, the ambiguous expression of the (first) gauge transformation $\delta \Psi _{1,-1} = Q \, \mu _{0,-1} + \eta \, \mu  _{0,0}$ provides larger and independent set of higher gauge parameters. 
In particular, the $(g+1)$-th gauge transformations take the form of $\delta _{g} \mu _{1-g,p} = Q \, \mu _{g,p} + \eta \, \mu _{g,p+1}$ as Berkovits theory \cite{Torii:2011zz, Kroyter:2012ni, Torii:2012nj, Iimori:2015aea}, and the $p$-label can run over all integer numbers. 
It requires many additional fields--antifields $\{ \Psi _{g,p} , ( \Psi _{g,p} )^{\ast } \} _{g\leq 1,p \in \mathbb{Z}}$ into the BV spectrum. 
As we showed in section 3.2, if we take a gauge variation $\delta \Psi = Q \, \eta \, \Lambda _{-1,0}$ using a \textit{large} gauge parameter $\Lambda _{-1,0} \in \cH _{\xi \eta \phi }$, the expression of the large gauge invariance is no longer ambiguous. 
Then, additional fields-antifields $\{ \Phi _{-1-g,p} , (\Phi _{-1-g,p})^{\ast } \}_{0 \leq p \leq g}$ are nothing but those of Berkovits theory. 
But it requires the BV formalism in the large Hilbert space \cite{Matsunaga:2018hlh}. 
In this sense, the additional fields--antifields arise from ambiguous expressions of the gauge invariances of superstring field theory based on the small Hilbert space, which is a result of the very trivial embedding (\ref{kinetic}) into the large Hilbert space. 

\vspace{1mm} 

This underling gauge structure of superstring field theory based on the small Hilbert space resembles that of the WZW-like formulation. 
Recently, it was shown that superstring field theory based on $\cH _{\beta \gamma }$ can be embedded into the WZW-like formulation unless taking their gauge reducibility into account \cite{Matsunaga:2016zsu, Erler}. 
In several example, it is known that classical actions based on $\cH _{\beta \gamma }$ can be obtain from WZW-like ones via field redefinitions reducing gauge symmetries \cite{Erler:2015rra, Erler:2015uba, Goto:2015hpa}, which anticipates corresponding canonical transformations. 
We may expect some exact relations of these at the level of their master actions. 
We end this paper with some remarks about it.

\subsection{On the general WZW-like formulation}

The general WZW-like formulation is a purely algebraic generalisation of the geometrical framework explained in section 1, in which the linear $\eta $-constraints on the states are extend to a nonlinear $\mathbf{C}$-constraints based on the homotopy algebra $\mathbf{C}$ whose linear part is $\eta$ \cite{Matsunaga:2016zsu ,Erler}. 
For simplicity, we take open superstring field theory as an example. 
Let $( \mathbf{C} , \mathbf{V} )$ be a mutually commutative pair of $A_{\infty }$ algebras: $\mathbf{C}$ is some nonlinear extension of $\eta$ and $\mathbf{V}$ is the string vertices $\{ Q , V_{g,n} \} _{g,n}$ given in section 1. 
Then, using a dynamical string field $\varphi $ of the theory, we consider a solution $A_{C} [ \varphi ]$ of the Maurer-Cartan equation for $\mathbf{C}$, 
\begin{align*}
\pi _{1} \mathbf{C} \frac{1}{1-A_{C} [\varphi ] } = 0 \, . 
\end{align*}
This $A_{C}[\varphi ]$ is a functional of the dynamical string field $\varphi $. 
Note that $\Psi \in \cH_{\beta \gamma }$ satisfy $\eta \Psi = 0$ and gives a trivial example for the case of $\mathbf{C} = \Eta $. 
Note also that when we take $\mathbf{C} = \Eta - {\bf m_{2}}$ and $\mathbf{V} = {\bf Q}$ ($m_{2}$ is Witten's star product), it reduces to the Berkovits theory \cite{Berkovits:1995ab}. 

When $\mathbf{D}$ is a derivation operator for $\mathbf{C}$, or more generally, an $A_{\infty }$ product $\mathbf{D}$ commuting with $\mathbf{C}$, one can define a functional $A_{\mathbf{D} }[\varphi ]$ such that 
\begin{align*}
(-)^{\mathbf{D}} \, \mathbf{D} \, \frac{1}{1-A_{C} [\varphi ] } = \mathbf{C} \, \frac{1}{1-A_{C} [\varphi ] } \otimes A_{\mathbf{D} } [\varphi ] \otimes \frac{1}{1-A_{C} [\varphi ] } \, . 
\end{align*} 
It is a generalisation of the relation $\partial _{t} ( e^{t \phi } d e^{-t \phi } ) = d ( e^{t \phi } \partial _{t} e^{-t \phi } ) + [ e^{t \phi } d e^{-t \phi } ,  e^{t \phi } \partial _{t} e^{-t \phi } ]$ satisfied by a pure-gauge state $e^{- t \phi } (d \, e^{t \phi })$ of Chern-Simons theory ($d$ is the exterior derivative and the product is the wedge product.).  
For a real parameter $t \in [ 0 , 1 ]$, its partial differential $\partial _{t}$ works as a derivation for $\mathbf{C}$. 
The variation $\delta $ of the field also satisfies the Liebniz rule for $\mathbf{C}$. 
Thus, one can take $\mathbf{D} = \partial _{t}$ or $\mathbf{D} = \delta $ for example. 
Because of mutual commutativity, one can also take $\mathbf{D} = \mathbf{V}$. 
Using these, the general WZW-like action $S_{\sf wzw}$ is given by 
\begin{align}
S_{\sf wzw} [\varphi ] = \int _{0}^{1} dt \, \La \, A_{\partial _{t} } [t \varphi ] \, , \, \pi _{1} \, \mathbf{V} \frac{1}{1-A_{C} [t \varphi ]} \, \Ra  \, . 
\end{align}
This $S_{\sf wzw} [\varphi ]$ gives a gauge invariant action for any $A_{\infty }$ pairs $( \mathbf{C} , \mathbf{V} )$. 
One can quickly get a proof by omitting one of the constraints in \cite{Matsunaga:2016zsu}. 
See also \cite{Erler} for detailed and pedagogical explanations about the general WZW-like action. 
The gauge transformations are 
\begin{align*}
A_{\delta } [\varphi ] = \pi _{1} \Big( \big[ \mathbf{C} , \bLambda _{C} \big] + \big[ \mathbf{V} , \bLambda \big] \Big) \frac{1}{1-A_{C} [\varphi ] } \, . 
\end{align*}
Here, $\bLambda _{C}$ and $\bLambda $ are gauge parameters. 
See \cite{Erler} or section 7 of \cite{Kajiura:2003ax} for the coalgebraic notation. 
In general, field redefinitions $\widehat{\bf U}$ drastically change the string vertices $\mathbf{V}$ in highly nontrivial manner. 
In terms of the $A_{\infty }$ pairs, it just gives a (weak) $A_{\infty }$ morphism between two $A_{\infty }$ pairs, $\widehat{\bf U} : ( \mathbf{C} , \mathbf{V}) \rightarrow ( \mathbf{C}' , \mathbf{V}' )$\,, which preserves the solutions of the Maurer-Cartan equations but may change the forms of the above functionals. 
Hence, this $S_{\sf wzw}[\varphi ]$ is covariant under string field redefinitions.\footnote{For type II theory, see \cite{Matsunaga:2016zsu}. 
One can consider heterotic theory by omitting one of the constraint $L_{\infty }$ algebras in type II theory or by replacing the pair of $A_{\infty }$ algebras of open superstring theory with that of $L_{\infty }$ algebras.}  
Therefore, as a gauge field theory, $S_{\sf wzw} [\varphi ]$ and its Batalin-Vilkovisky master action $S_{\sf BV}$ will capture very general properties of superstrings, which is also supported by our results obtained in section 2.   

Unfortunately, we do not have enough understandings about the most general form of $S_{\sf BV}$ yet. 
In the rest, we consider a slightly generalised version of our analysis, the same pair $( \Eta , \mathbf{V} )$ but large fields $\varphi \in \cH_{\xi \eta \phi }$, which is the second simplest but first nontrivial example of $S_{\sf wzw}[\varphi ]$. 

\sub{Example: Large $A_{\infty }$ open superstring field theory}

We write $\Phi _{0,0}$ for a dynamical string field, which lives in the state space of $\xi \eta \phi $-system, the large Hilbert space $\cH_{\xi \eta \phi }$. 
Thus, there is no constraint on the string field $\Phi _{0,0}$\,. 
In this set up, we find $A_{C} [\Phi ] = \eta \, \Phi _{0,0}$\,, and the classical action is given by 
\begin{align*}
S _{\sf A} [ \Phi ] & = \frac{1}{2} \la \Phi _{0,0} , \, Q \, \eta \, \Phi _{0,0} \ra  + \sum_{n>1} \frac{1}{n+1} \la \xi \, \eta \, \Phi  _{0,0} , \, M_{n} \big( \overbrace{ \eta \, \Phi _{0,0} , \dots , \eta \, \Phi  _{0,0} }^{n} \big) \ra \, , 
\end{align*}
which takes the same form as (\ref{action open}) except for the dynamical string field, $S_{\sf A} [\Phi ] = S[\eta \Phi ]$. 
This theory has large gauge invariances generated by two gauge generators: 
\begin{align*}
\delta \Phi _{0,0} = \eta \, \Lambda _{-1,1} + \sum_{\rm cyclic} \sum_{n=0}^{\infty } M_{n+1} \big( \overbrace{ \eta \, \Phi _{0,0} , \dots , \eta \, \Phi _{0,0} }^{n} , \Lambda_{-1,0} \big) \, , 
\end{align*}
where we write $\Lambda _{-g,p}$ for a gauge parameter field living in $\cH_{\xi \eta \gamma }$. 
Since its kinetic term is that of the Berkovits theory, it is infinitely reducible gauge theory \cite{Torii:2011zz, Kroyter:2012ni}. 
The gauge invariance of the kinetic term is given by $\delta \Phi _{0,0} = \eta \, \Lambda _{-1,1} + Q \, \Lambda _{-1,0}$. 
Using the nilpotency of $(Q)^{2} = (\eta )^{2} =0$ and the graded commutation relation $\eta \, Q + Q \, \eta  = 0$, we find the $g$-th gauge transformations for the $(g-1)$-th gauge transformations $\delta _{g} \Lambda _{-g,p} = \eta \, \Lambda _{-g-1,p+1} + Q \, \Lambda _{-g-1,p}$ satisfying $\delta _{g} ( \delta _{g-1} \Lambda _{1-g,p} ) = 0$ for $0 \leq p \leq g$. 
Since these gauge parameter fields turn into ghosts $\{ \Phi _{g,p} \}_{g\leq -1, 0 \leq p \leq |g| }$ and they lead antighosts, the minimal set of the fields--antifields is given by 
\begin{align}
\label{symbl} 
\big{\{ } \Phi _{-g,p} , \Phi _{2+g,-1-p}^{\ast } \big{\} } _{0 \leq g, 0 \leq p \leq g} \subset \cH _{\xi \eta \phi } \, . 
\end{align} 
Note that there is no constraints on the BV spectrum. 
The free master action takes the same form as (\ref{free master action}) except for the BV spectrum. 
See \cite{Torii:2011zz, Kroyter:2012ni, Torii:2012nj, Iimori:2015aea} for details. 

\sub{Large master action for interacting theory} 

As we showed in section 3, we can construct the master action in the canonical form, which will be a canonical-transformed version of the unclear original form of $S_{\sf BV}$. 
Unfortunately, we do not have clear understanding about the most general form of $S_{\sf BV}$. 
However, in this case, one can find a more enlarged form of $S_{\sf BV}$. 
See also \cite{Matsunaga:2018hlh}. 
We set 
\begin{align*} 
\varphi \equiv \sum_{g\geq 0} \sum_{p=0}^{g} \Big[ \Phi _{-g,p} + \Phi ^{\ast }_{2+g,-1-p} \Big] \, .
\end{align*} 
Note that although this $\varphi $ is given by using the same symbols as (\ref{symbl}), these fields--antifields $\{ \Phi _{-g,p} , (\Phi _{-g,p})^{\ast } \} _{0 \leq p \leq g}$ should be regarded as some canonical transformed ones from (\ref{symbl}). 
Let $\epsilon $ be an operator counting the grading of the state: $\epsilon \, \Phi = (-)^{\epsilon [ \Phi ] } \Phi = (-)^{|\Phi |} \Phi$. 
Then, we find that the following $S_{\sf BV}$ satisfies the master equation, 
\begin{align*}
S_{\sf BV} [ \varphi ] 
& = \int _{0}^{1} dt \, \La \, \varphi - \epsilon \, \eta \, \varphi \, , \, \pi _{1} (\Eta + \bM ) \frac{1}{1- t \, \varphi - t \, \epsilon \, \eta \, \varphi } \, \Ra  
\no 
& = \frac{1}{2} \la \, \varphi \, , \, \eta \, \varphi \, \ra 
+ \sum_{n=1}^{\infty } \frac{1}{n+1} \La \, \varphi - \epsilon \, \eta \, \varphi \, , \, M_{n} \big( \overbrace{ \varphi + \epsilon \, \eta \, \varphi \, , \dots , \varphi + \epsilon \, \eta \, \varphi }^{n} \big) \, \Ra \, . 
\end{align*}
Since the space-time ghost number of $S_{\sf BV} [\varphi ]$ equals to zero, for which we write $s(S_{\sf BV} [\varphi])=0$, its total degree is also zero: $\epsilon ( S_{\sf BV} [\varphi ] ) = 0$\,. 
The variation of $S_{\sf BV}$ is given by 
\begin{align*}
\delta S_{\sf bv} [\varphi ] 
& = \La \, \delta \varphi \, , \, \pi _{1} \big( \Eta + \bM - \Eta \, \bM \big) \frac{1}{1- \varphi - \epsilon \, \eta \, \varphi } \, \Ra  
\no 
& = \la \, \delta \varphi \, , \, \eta \, \varphi \, \ra 
+ \sum_{n=1}^{\infty } \La \, \delta \varphi \, , \, M_{n} \big( \overbrace{ \varphi + \epsilon \, \eta \, \varphi \, , \dots }^{n} \big) 
- \eta \, M_{n} \big( \overbrace{ \varphi + \epsilon \, \eta \, \varphi \, , \dots }^{n} \big) \, \Ra \, . 
\end{align*}
We thus find that the gauge invariance of $S_{\sf BV} [\varphi ]$, or the BV-BRST transformations of the BV master action, is given by 
\begin{align*}
\delta \varphi = \eta \, \varphi + \pi_{1} \big[ \bM - \Eta \, \bM \big] \frac{1}{1-\varphi - \epsilon \, \eta \, \varphi } \, . 
\end{align*}
We write $(s; g,p)$ of $|^{s;}_{g,p}$ for the projection onto the space-time ghost number $s$\, world-sheet ghost number $g$\,, and picture number $p$ state. 
Then, one can express the BV-BRST transformations $\delta _{\rm BV} \Phi _{-g,p}$ and $\delta _{\rm BV} \Phi _{1+g,-p}^{\ast }$ as follows 
\begin{align*}
\frac{\partial _{l} S_{\sf BV} [\varphi ]}{\partial \Phi ^{\ast }_{1+g,-p} } 
& = \eta \, \Phi _{-g,p} 
+ \sum_{n=1}^{\infty } \Big[ 
M_{n} \big( \overbrace{ \varphi + \eta \, \varphi \, , \dots }^{n} \big) 
- \eta \, M_{n} \big( \overbrace{ \varphi + \eta \, \varphi \, , \dots }^{n} \big)  
\Big] ^{g;}_{1-g,p-1} \, , 
\\ 
\frac{\partial _{r} S_{\sf BV} [\varphi ]}{\partial \Phi _{-g,p} } 
& = \eta \, \Phi ^{\ast }_{1+g,-p} 
+ \sum_{n=1}^{\infty } \Big[ 
M_{n} \big( \overbrace{ \varphi + \eta \, \varphi \, , \dots }^{n} \big) 
- \eta \, M_{n} \big( \overbrace{ \varphi + \eta \, \varphi \, , \dots }^{n} \big)  
\Big] ^{-g;}_{2+g,-1-p} \, . 
\end{align*}
These derivatives provide the classical BV master equation. 
Recall that $A_{\infty }$ relations imply 
\begin{align*}
\sum_{i=1}^{\infty } \sum_{j=1}^{\infty } 
\La M_{i} ( A , ... , A ) , M_{j} ( A , \dots , A ) \Ra 
& = \sum_{n=1}^{\infty } \sum_{l=0}^{n-1} 
\La M_{l+1} ( A , ... , A ) , M_{n-l} ( A , \dots , A ) \Ra 
\no & \hspace{-45mm} 
= \sum_{n=1}^{\infty } \frac{1}{n+1} \sum_{l=0}^{n-1} \big[ (l+1) + (n-l) \big] 
\La M_{l+1} ( A , ... , A ) , M_{n-l} ( A , \dots , A ) \Ra 
\no & \hspace{-50mm} 
= \sum_{n=1}^{\infty } \frac{2}{n+1} \sum_{l=0}^{n-1} \Big[ \sum _{k=0}^{l} 
\La A , M_{l+1} ( \overbrace{A , ... }^{k}, M_{n-l} ( A , \dots , A ) , \overbrace{... , A}^{l-k} ) \Ra \Big] 
= 0 \, . 
\end{align*}
Because of these $A_{\infty }$ relations and $(\eta )^{2} = 0$\,, we quickly find 
\begin{align*}
\frac{1}{2} \Big{\{ } \, S_{\sf BV} [\varphi ] \, , \, S_{\sf BV} [\varphi ] \, \Big{\} } 
& = \sum_{m} \La \eta \, \varphi , \, M_{m} ( \varphi + \epsilon \, \eta \, \varphi , ... ) \Ra 
+ \sum_{n} \La M_{n} ( \varphi + \epsilon \, \eta \, \varphi , ... ) , \,   \eta \, \varphi \Ra 
= 0 \, , 
\end{align*}
which is the mutual commutativity of $\Eta $ and $\bM $\,. 
Note that the antibracket $\{ S_{\sf BV} , S_{\sf BV} \} $ has space-time ghost number one, $s (\{ S_{\sf BV} , S_{\sf BV} \} )=1$\,, and thus its total degree is one, $\epsilon (\{ S_{\sf BV} , S_{\sf BV} \} ) =1$\,. 
Hence, in this case, $\la A , B \ra = - \la B , A \ra $ holds for any states $A, B$ satisfying $s(A) + s(B) =1$\,. 

\vspace{2mm}

Off course, one may be able to construct a more enlarged form of the master action including not only $\eta$ but also many $\bM$. 
We would like to emphasise that a natural perturbative construction based on the antifield number expansion anticipates such a larger solution. 
Thus, the above $S_{\sf BV}$ will be also canonical-transformed one. 
These unknown but interesting feature of the gauge invariances may be understood by canonical transformations as we shown in section 2.
We expect that our results gives a first step to obtain clear insights into the gauge structure and field theoretical properties of superstrings. 
See \cite{Matsunaga:2018hlh} for other types of master actions.

\section*{Acknowledgements} 

The author would like to thank Theodore Erler, Hiroshi Kunitomo, Martin Schnabl. 
The author also thank Hiroshige Kajiura and Jim Stasheff. 
This research has been supported by the Grant Agency of the Czech Republic, under the grant P201/12/G028. 

The author thanks Nathan Berkovits for email.

\appendix 

\section{On the string fields--antifields} 

In this appendix A.1, after explaining some properties of the BPZ inner product and its basis, we give a string field representation of the BV antibracket. 
In appendix A.2, we first give the free BV master action for the exactly small theory. 
Then, we show how the BV spectrum is changed by relaxing the $\eta $-constraints on higher ghost-fields--antifields of the exactlly small theory.  

\subsection{String antibracket} 

Let $\{ \Psi _{g,p} , ( \Psi _{g,p} )^{\ast } \} _{g,p} \subset \cH _{\xi \eta \phi }$ be the minimal set of fields--antifields in superstring field theory. 
While the $g$-label of $\Psi _{g,p}$ denotes that it corresponds to the $g$-th ghost fields, the $p$-label of $\Psi _{g,p}$ distinguishes their difference at the same $g$-label. 
Then, the antibracket $\{ F ,G \}$ of two functions $F = F [\Psi , \Psi ^{\ast }]$ and $G = G[\Psi , \Psi ^{\ast } ]$ can be presented by  
\begin{align}
\label{string field rep}
\big{\{ }  F , \, G \big{\} } = \sum_{g} \sum_{p} \bigg[ 
\LA \frac{\partial _{r} F}{\partial \Psi _{g,p}} , \, \frac{\partial _{l} G}{\partial \Psi _{g,p}^{\ast }} \RA  
- \LA \frac{\partial _{r} F}{\partial \Psi _{g,p}^{\ast }} , \, \frac{\partial _{l} G}{\partial \Psi _{g,p}} \RA \bigg]  \, , 
\end{align}
where $\langle A, B \rangle $ is the large BPZ inner product of two string fields $A$ and $B$. 
One can quickly derive this string field representation of the BV antibracket (\ref{string field rep}) by direct computations of usual BV antibracket under two assumptions about string fields and their functions. 

\sub{Graded symplectic BPZ product} 

The large BPZ inner product $\langle A , \, B \rangle _{\sf bpz}$ takes nonzero value if and only if the sum $g$ of the inputs' world-sheet ghost numbers and the sum $p$ of the inputs' picture numbers equals to the appropriate value\footnote{We consider the $(c_{0}-\tilde{c}_{0})$-inserted one for closed string field theory, in which all string fields $\Psi $ satisfy $(b_{0}-\tilde{b}_{0}) \Psi = 0$ and $(L_{0} - \tilde{L}_{0}) \Psi = 0$.}: $(g,p)=(2,-1)$ for open strings, $(g,p)=(4,-1)$ for heterotic strings, and $(g,p,\tilde{p})=(3,-1,-1)$ for type II strings. 
Its Grassmann parity depends on the theory. 
In this paper, we introduce an appropriate grading, so-called degree, and consider suspended versions, in which string fields have degree $0$ and the string vertices have degree $1$ (see \cite{Gaberdiel:1997ia, Erler:2014eba}.). 
For open superstring field theory based on $\Psi \in \cH_{\beta \gamma }$, the degree of states are defined by ``space-time ghost number $+$ world-ghost number $-1$''. 
For heterotic and type II theories based on $\Psi \in \cH_{\beta \gamma }$, the degree of states are defined by ``space-time ghost number $+$ world-ghost number $-2$''.  
For example, our large BPZ inner product $\langle A , B \rangle $ is given by 
\begin{align*}
\la A , \, B \ra \equiv (-)^{{\sf G} ( A )} \la A , \, B \ra _{\sf bpz} \,, 
\end{align*}
where ${\sf G}(A)$ denotes the Grassmann parity of the state $A$. 
Then, the large BPZ inner product becomes symplectic, all BV fields $\Psi \in \cH _{\beta \gamma }$ have degree $0$, and all antifields $ (\Psi )^{\ast } \in \cH _{\xi \eta \phi }$ have degree $-1$. 
We write $\epsilon [\Psi ]$ for the (total) grading of $\Psi $, its (total) degree.

In string field theory, string fields $\{ \Psi _{g,p} \} _{g,p}$ consist of a set of space-time fields $\{ A_{g,p} \} _{g,p}$ and a set of world-sheet basis $\{ \cZ _{g,p} \} _{g,p}$, for which we write $\Psi _{s;g,p} \equiv A_{s,p} \, | \cZ _{g,p} \rangle$\,. 
As well as the world-sheet basis $\cZ _{g,p}$, the space-time field $A_{s,p}$ also has its grading $\epsilon [A_{s,p}]$ which is equal to its space-time ghost number $s[A_{s,p}]=s$. 
We thus find $\epsilon [ \Psi _{s;g,p}] = \epsilon [A_{s,p} ] + \epsilon [ \cZ _{g,p} ]$ and 
\begin{align*}
\la \Psi _{s;g,p} , \, \Psi _{t;h,q} \ra = - (-)^{{\epsilon } [ \Psi _{s;g,p} ] \epsilon [ \Psi _{t;h,q}] } \la \Psi_{t;h,q} , \, \Psi _{s; g,p} \ra \, . 
\end{align*}
We write $\epsilon [ \omega ]$ for the grading of the symplectic BPZ product: $\epsilon [ \langle A,B \rangle ] = \epsilon [ \omega ] + \epsilon [ A ] + \epsilon [B]$. 
In this paper, our computations are based on the following defining relations 
\begin{align*}
\la A_{s} \, \cZ _{g,p} , \, A_{t} \, \cZ _{h,q} \ra 
= (-)^{{\epsilon } [A_{t}] {\epsilon } [ \cZ_{h,q} ] } \la A_{s} \, \cZ _{g,p} , \, \cZ _{h,q} \ra \, A_{t}  
= (-)^{{\epsilon } [\omega ] {\epsilon } [A_{s} ] } A_{s} \, \la \cZ _{g,p} , \, A_{t} \, \cZ _{h,q} \ra \, . 
\end{align*}

\sub{Batalin-Vilkovisky antibracket} 

We write $\{ A_{g,p} , (A_{g,p})^{\ast } \} _{g,p}$ for the minimal set of space-time fields--antifields in usual gauge field theory. 
The $g$-label of $A_{g,p}$ denotes that $A_{g,p}$ corresponds to the $g$-th gauge reducibility, namely, its space-time ghost number, for which we write $s [ A_{g,p}]= g$\,. 
Then, by construction, corresponding antighost $(A_{g,p})^{\ast }$ has space-time ghost number $-(g+1)$, for which we write $s [ (A_{g,p})^{\ast } ] = -(g+1)$. 
Let $F= F[A,A^{\ast }]$ and $G = G[A,A^{\ast }]$ be functions of these space-time fields--antifields. 
In the Batalin-Vilkovisky formalism, using these space-time ghost--antighost fields, the BV antibracket $\{ F , G \}$ is given by 
\begin{align} 
\label{usual antibracket}
\{ F , \, G \} = \sum_{g} \sum_{p} \left( \frac{\partial _{r} F}{\partial A_{g,p}} \frac{\partial _{l} G }{\partial A^{\ast }_{g,p} } - \frac{\partial _{r} F}{\partial A^{\ast }_{g,p}} \frac{\partial _{l} G }{\partial A_{g,p} } \right) \, . 
\end{align}
The antibracket has space-time ghost number $+1$ and $s[ \{ F ,G \} ]=s[F]+s[G] +1$ holds. 
Recall that computations of the antibracket is based on the following expression of the variation, 
\begin{align*}
\delta F [ A , A^{\ast } ] = 
\sum_{g,p} \bigg[ 
\frac{\partial _{r} F}{\partial A_{g,p}} \, \delta A_{g,p} + \frac{\partial _{r} F}{\partial A^{\ast }_{g,p}} \, \delta A^{\ast }_{g,p} 
\bigg] 
= \sum_{g,p} \bigg[ 
\delta A_{g,p}  \, \frac{\partial _{l} F }{\partial A_{g,p} } 
+ \delta A^{\ast }_{g,p} \, \frac{\partial _{l} F }{\partial A^{\ast }_{g,p} } 
\bigg] \, . 
\end{align*}

\sub{String field representation} 

We would like to obtain a string field representation of the BV antibracket. 
Let $\lambda _{1-g,p}$ be the $g$-th gauge parameter fields, which consists of sets of space-time gauge parameter fields $\{ B_{g,p} \} _{p}$ and world-sheet basis $\{ \cZ _{1-g,p} \}_{p}$. 
Then, its ghost field $\Psi _{1-g,p}$ is obtained by replacing $\cB _{g,p}$ of $\lambda _{g,p}$ by corresponding $g$-th space-time ghost field $A_{g,p}$, namely, $\Psi _{1-g,p} = A_{g,p} \, | \cZ _{1-g,p} \rangle$\,.  
Hence, the field $\Psi _{1-g,p}$ has space-time ghost number $g$, world-sheet ghost number $1-g$, and Grassmann parity $1$, for which we write $s[\Psi _{g,p} ] = g$, ${\rm gh}[\Psi _{g,p} ] =1-g$, and ${\sf G} [\Psi _{g,p} ] =1$ respectively. 
Then, the antifield $(\Psi _{1-g,p} )^{\ast }$ consists of sets of the $g$-th space-time `antighost' fields $\{ A_{g,p}^{\ast } \} _{p}$ having space-time ghost number $-(g+1)$ and the world-sheet basis $\{ \cZ ^{\ast }_{1-g , p} \}_{p} $ satisfying $\langle \cZ_{g,p} , \cZ _{g,p}^{\ast } \rangle \not= 0$. 
Therefore, as we saw in section 2, we have $\Psi _{1-g,p}^{\ast } = A^{\ast }_{g} \, | \cZ _{1+g, 0} \rangle $ for open superstring field theory. 
Note that all fields $\{ \Psi _{g,p} \} _{g,p}$ have degree zero and all antifields $\{ \Psi ^{\ast }_{g,p} \} _{g,p}$ have degree $-1$, for which we write $\epsilon [\Psi _{g,p}] = 0$ and $\epsilon [\Psi _{g,p}^{\ast } ] = -1$. 
Note also that $s[(\Psi _{g,p} )^{\ast } ]= - (g+1)$. 

Let $F=F[\Psi , \Psi ^{\ast } ]$ be a function of given minimal set of fields--antifields $\{ \Psi _{g,p} , (\Psi _{g,p})^{\ast } \} _{g,p}$. 
In string field theory, any $\mathbb{C}$-value function of string fields is written by using the BPZ inner product. 
Then, we assume that the variations of fields--antifields are given by 
\begin{align*}
\delta \Psi _{g,p} \equiv \delta A_{g,p} \, | \cZ _{g,p} \rangle \, ,
\hspace{5mm}  
\delta \Psi ^{\ast }_{g,p} \equiv \delta A^{\ast }_{g,p} \, | \cZ ^{\ast }_{g,p} \rangle \, .
\end{align*} 
Likewise, we assume that the variation of any $\mathbb{C}$-value functional $F [ \Psi , \Psi ^{\ast }]$ is given by 
\begin{align*}
\delta F [ \Psi , \Psi ^{\ast }] & \equiv 
\sum_{g,p} \hspace{-0.3mm} \bigg[ \hspace{-0.3mm} 
\La \delta \Psi _{g,p} , \frac{\partial _{l} F}{\partial \Psi _{g,p}} \Ra 
\hspace{-0.3mm} + \hspace{-0.3mm} 
\La \delta \Psi _{g,p}^{\ast } , \frac{\partial _{l} F}{\partial \Psi _{g,p}^{\ast }} \Ra \hspace{-0.3mm} \bigg] 
= \sum_{g,p} \hspace{-0.3mm} \bigg[ \hspace{-0.3mm}
\La \frac{\partial _{r} F}{\partial \psi _{g,p}} , \, \delta \Psi _{g,p} \Ra 
\hspace{-0.3mm} + \hspace{-0.3mm} 
\La  \frac{\partial _{r} F}{\partial \Psi _{g,p}^{\ast }} , \, \delta \Psi _{g,p}^{\ast } \Ra \hspace{-0.3mm} \bigg] . 
\end{align*}
On the basis of these string field representations of the variations, the BV antibracket (\ref{usual antibracket}) is written into its string field representation (\ref{string field rep}). 
The string field derivatives are defined in the same manner. 
Then, we have to pay attention to the grading of the inner product.

\subsection{Constraints on BV spectrums} 

We consider to relax the restrictions on the gauge parameters step by step. 
Then, the BV spectrums are enlarged, and the master actions has larger gauge invariances. 
For brevity, in this appendix, we consider open superstring fields as an example. 

\sub{The exactly small theory}

First, let us consider to restrict the dynamical string field $\Psi _{1,-1}$ and the gauge parameter field $\lambda _{0,-1}$ onto the small Hilbert space $\cH _{\beta \gamma }$. 
Using the small BPZ inner product, the gauge invariant action is given by 
\begin{align}
\label{exact small}
S_{\sf s} = \frac{1}{2} \lla \Psi _{1,-1} , \, Q \, \Psi _{1,-1} \rra \, , \hspace{5mm} 
\eta \, \Psi _{1,-1} = 0 \, . 
\end{align}
Because of the constraints, this theory is gauge invariant under the (first) gauge transformation 
\begin{align*}
\delta \, \Psi _{1,-1} = Q \, \lambda _{0,-1}  \, , \hspace{5mm } \eta \, \lambda _{0,-1} = 0 \, .
\end{align*}
It yields the gauge reducibility $\delta _{n+1} ( \delta _{n} \lambda _{-n,-1} ) = 0$ under the $n$-th gauge transformation for the $(n-1)$-th gauge transformation 
\begin{align*}
\delta _{n} \lambda _{-n,-1} = Q \, \lambda _{-(n+1),-1} \, . 
\end{align*} 
While $\Psi _{1,-1}$ and $\lambda _{0,-1}$ must satisfy the constraint equations $\eta \, \Psi _{1,-1}= 0$ and $\eta \, \lambda _{0,-1}= 0$\,, the other higher gauge parameters  $\{ \lambda _{-g,-1} \} _{g>1}$ do not have to satisfy any constraint equation. 
However, when we use the small BPZ inner product to obtain the BV master action, we have to impose the same constraint equations on the all higher gauge parameters, 
\begin{align*} 
 \eta \, \lambda _{-n,-1} = 0 \, , \hspace{5mm}    ( n \in \mathbb{N} ) \, . 
\end{align*} 

Then, the set of gauge parameters $\{ \lambda _{g,-1} \} _{g\leq 0}$ appears in the theory. 
It implies that the minimal set of fields--antifields is given by 
\begin{align*}
\big{\{ } \Psi _{1,-1} , \, \Psi _{-g,-1} \big{\} }_{g\geq0} \subset \cH _{\beta \gamma } \, , \hspace{5mm} 
\big{\{ } \Psi _{2+g,-1}^{\ast } \equiv (\Psi _{1-g,-1} )^{\ast } \big{\} }_{g \geq 0} \subset \cH  _{\beta \gamma } \, . 
\end{align*}
We write $(\Psi _{g,p} )^{\ast }$ for the antifield of $\Psi _{g,p}$, whose ghost and picture numbers is determined via the BPZ inner product of  the theory as $\slla (\Psi _{g,p} )^{\ast } , \Psi _{g,p} \srra \not= 0$. 
We thus write $\Psi ^{\ast }_{2+g , -1}$ for the antifield of $(\Psi _{1-g,-1})^{\ast }$\,, namely $\Psi ^{\ast }_{2+g, -1} \equiv ( \Psi _{1-g , -1-p})^{\ast }$\,.
These fields and antifields must satisfy the constraint equations  
\begin{align*}
\eta \, \Psi _{1-g , -1} = 0 \, , \hspace{5mm} 
\eta \, \Psi ^{\ast }_{2+g , -1} = 0 \, , \hspace{5mm} 
( g \in \{ 0 \} \cup \mathbb{N} ) \, . 
\end{align*}
Note that for each field $\Psi $ or each antifield $\Psi ^{\ast }$, the sum of the space-time and world-sheet ghost numbers is just $1$. 
In other words, the minimal set consists of the degree $0$ states: $\epsilon [ \Psi _{1-g,-1} ] = 0$ and $\epsilon [ \Psi _{2+g,-1}^{\ast } ] = 0$\,. 
Using this minimal set, one can construct the master action, 
\begin{align*}
S_{\sf bv, s} [ \Psi , \Psi ^{\ast } ] = \frac{1}{2} \lla \Psi _{1,-1} , \, Q \, \Psi _{1,-1} \rra 
+ \sum_{g=1}^{\infty } \lla \Psi ^{\ast }_{2+(g-1),-1} , \, Q \, \Psi _{1-g,-1} \rra \, . 
\end{align*}
Since $\frac{\partial _{r} S_{\sf bv, s} [\Psi , \Psi ^{\ast }] }{\partial \Psi _{1-g,-1}} = Q \, \Psi ^{\ast }_{1+g,-1}$ and $ \frac{\partial _{r} S_{\sf bv, s} [\Psi , \Psi ^{\ast }] }{\partial \Psi ^{\ast }_{2+g,-1}} = Q \, \Psi _{-g,-1}$\,, it solves the master equation 
\begin{align*}
\frac{1}{2} \big{\{ } S_{\sf bv, s}  , \, S_{\sf bv, s}  \big{\} } = \sum_{g=0}^{\infty } \lla Q \, \Psi ^{\ast }_{1+g,-1} , \, Q \, \Psi _{-g,-1} \rra = 0 \,. 
\end{align*}

\sub{Very trivial embedding into the large Hilbert space} 

We can re-express the action (\ref{exact small}) using the large BPZ inner product 
\begin{align*}
S_{\sf l} = \frac{1}{2} \la \xi \, \Psi _{1,-1} , \, Q \, \Psi _{1,-1} \ra \,  \hspace{5mm } \eta \, \Psi _{1,-1} = 0 \, . 
\end{align*}
Iff we impose $\eta \, \lambda _{0,-1} = 0$, it has the same gauge invariance $\delta \, \Psi _{1,-1} = Q \, \lambda _{0,-1}$ with $\eta \, \lambda _{0,-1}= 0$\,. 
As well as $\lambda _{0,-1}$, the other higher gauge parameters $\{ \lambda _{-g} \} _{g>1}$ do not have to satisfy any constraint equation. 
However, if we restrict these on $\cH _{\beta \gamma }$, namely $\eta \, \lambda _{-n,-1}=0$ for $n\geq 0$, this theory has the same gauge reducibility $\delta _{n+1} ( \delta _{n} \lambda _{-n} ) = 0$ under the $n$-th gauge transformation for the $(n-1)$-th gauge transformation $\delta _{n} \lambda _{-n} = Q \, \lambda _{-(n+1)}$\,. 

If we consider to construct the master action $S_{\sf bv, l}$ based on the large BPZ inner product, one can slightly enlarge the minimal set. 
All antifields $(\Psi _{g,-1})^{\ast }$ can live in the large Hilbert space $\cH _{\xi \eta \phi }$ because of $\langle (\Psi _{g,-1})^{\ast } , \Psi _{g,-1} \rangle \not= 0$. 
Then, the minimal set of fields--antifields is given by 
\begin{align*}
\big{\{ } \Psi _{1,-1} , \, \Psi _{-g,-1} \big{\} }_{g\geq0} \subset \cH _{\beta \gamma } \, , \hspace{5mm} 
\big{\{ } \Phi _{1+g,0}^{\ast } \equiv (\Psi _{1-g,-1} )^{\ast } \big{\} }_{g \geq 0} \subset \cH  _{\xi \eta \phi } \, . 
\end{align*}
We write $\Phi ^{\ast }_{1+g , 0}$ for the antifield of $\Psi _{1-g,-1}$\,, namely $\Phi ^{\ast }_{1+g, 0} \equiv (\Psi _{1-g , -1})^{\ast } $\,.
In this case, while all fields $\{ \Psi _{1-g} \} _{g\geq 0} = \{ \Psi _{1,-1}, \Psi _{0,-1} , ... , \Psi _{1-g,-1}, ... \}$ must satisfy the constraint equations, 
\begin{align*}
\eta \, \Psi _{1-g,-1} = 0 \, , \hspace{5mm} (g \geq 0) \, ,
\end{align*}
there is no constraint equations on the antifields. 
As we will see, essentially, the above constraints are too strong, and weaker constraints, $\eta \, Q \, \Psi _{-g,-1} = 0$ for $g \geq 0$, are sufficient for the master equation. 
One can find that the master action is given by 
\begin{align*}
S_{\sf bv, l} [ \Psi , \Phi ^{\ast } ; \cL ] & = \frac{1}{2} \la \xi \, \Psi _{1,-1} , \, Q \, \Psi _{1,-1} \ra 
+ \sum_{g=1}^{\infty } \la \Phi ^{\ast }_{g,0} , \, Q \, \Psi _{1-g,-1} \ra \, 
+ \sum_{g=0}^{\infty } \la \cL _{g,1} , \, \eta \, \Psi _{1-g,-1} \ra \, ,
\end{align*}
where we introduced Lagrange multipliers $\{ \cL _{g,1} \}_{g\geq 0}$ to see the role of the constraints. 
Since $\frac{\partial _{r} S_{\sf bv, l}}{\partial \Psi _{1,-1}} = \xi \,Q \, \Psi _{1,-1} + \eta \, \cL _{0,1}$\,, $\frac{\partial _{r} S_{\sf bv, l}}{\partial \Psi _{1-g,-1}} = Q \, \Phi ^{\ast }_{g,0} + \eta \, \cL _{g,1}$\,, and $\frac{\partial _{r} S_{\sf bv, l} }{\partial \Phi ^{\ast }_{1+g,0}} = Q \, \Psi _{-g,-1}$\,, after integrating out all Lagrange multipliers, we find 
\begin{align*}
\frac{1}{2} \big{ \{ }  S_{\sf bv, l} , \, S_{\sf bv, l}  \big{ \} } = 
\la Q \, \Psi _{0,-1} , \, \xi \, Q \, \Psi _{1,-1} \ra 
+ \sum_{g\geq 0} \la Q \, \Psi _{-g,-1} , \, \eta \, \cL _{g,1} \ra = 0 \, . 
\end{align*}
We write $\Psi /\Phi ^{\ast }$ for the sum of all fields/antifields. 
After inposing the constraints, we can rewrite the master action into the following form 
\begin{align*}
S_{\sf bv,l} [ \Psi , \Phi ^{\ast } ] 
& = \frac{1}{2} \la \xi \, ( \Psi + \eta \, \Phi ^{\ast } ) , \, Q \, ( \Psi + \eta \, \Phi ^{\ast } ) \ra  
= \frac{1}{2} \la \eta \, ( \xi \, \Psi + \Phi ^{\ast } ) , \, Q \, ( \xi \, \Psi + \Phi ^{\ast } ) \ra  
\, .
\end{align*}
Therefore, one can reduce it to $S_{\sf bv , s}[\Psi , \Psi ^{\ast }]$ by using gauge-fixing fermion providing $\eta \, \Phi ^{\ast } \equiv \Psi ^{\ast }$, or one may be able to regard it as some reduced version of larger master action.

\sub{Removing restrictions} 

What happens if we relax the above constraints on higher gauge parameters $\{ \lambda _{-g,-1} \} _{g}$\,? 
Let us consider to remove the constraints on the $(g \geq n)$ higher gauge parameters. 
Namely, we introduce two types of gauge parameters: $\{ \lambda _{-g ,-1} \} _{g=0}^{n-1} \subset \cH _{\beta \gamma }$ and $\{ \Lambda _{-g,-1} \} _{g=n}^{\infty } \subset \cH _{\xi \eta \phi }$\,. 
Then, $n$ gauge parameters $\lambda _{1-g,-1}$ satisfy the constraints 
\begin{align*}
\eta \, \lambda _{1-g,-1} = 0 \, , \hspace{5mm} (g=1 , \dots , n) \, ,
\end{align*}
and infinite number of gauge parameters $\{ \Lambda _{1-g,-1} \} _{g=n}^{\infty }$ are constraint free. 
Because of $\eta \, \lambda _{-g,-1} = 0$\,, we have 
\begin{align*}
\delta _{g} \lambda _{1-g,-1} = Q \, \lambda _{-g,-1} \, , \hspace{5mm} (g=1 , \dots , n-1) \, ,
\end{align*}
which preserves the $(n-1)$-th gauge transformation and constraint. 
If we require no restriction on the gauge variation $\delta _{n} \lambda _{1-n,-1}$, the following gauge reducibility arises 
\begin{align*}
\delta _{n} \lambda _{1-n,-1} = Q \, \Lambda _{-n,-1} \, , \hspace{5mm} 
\delta _{g} \Lambda _{1-g,-1} = Q \, \Lambda _{-g,-1} \, , \hspace{5mm} (g > n ) \, . 
\end{align*}
However, if we require $\delta _{n} \lambda _{1-n,-1} \in \cH _{\beta \gamma }$ as $\lambda _{1-n,-1} \in \cH _{\beta \gamma }$, we obtain the slightly different gauge reducibility. 
First, the $n$-th gauge parameter $\Lambda _{1-n,-1}$ cannot live in $\cH _{\xi \eta \phi }$ and must belong to ${\rm Ker}[Q\eta ]$ (or ${\rm Ker}[\eta ] \cup {\rm Ker} [Q]$). 
We thus write $\mu _{-n,-1} \equiv \frac{1}{2} \Lambda _{-n,-1}$\,. 
Because of $\mu _{-n,-1} \in {\rm Ker}[Q\eta ]$, there exist auxiliary gauge parameters $\{ \mu _{-n,p} \} _{p \in \mathbb{Z} \setminus \{ -1 \} } \subset {\rm Ker}[Q\eta ]$ such that 
\begin{align*} 
Q \, \mu _{-n,-1} = \eta \, \mu _{-n,0} \, , \hspace{5mm} 
\eta \, \mu _{-n,p} +Q \, \mu _{-n,p-1} = 0 \, , \hspace{5mm} (p \leq -1 , \, 1 \leq p ) \, .   
\end{align*}
These give the constraint equations on the $n$-th (auxiliary) gauge parameters $\{ \mu _{-n,p} \} _{p \in \mathbb{Z}}$, and thus these gauge parameters are dependent each other. 
Using them, we find $\delta _{n} \lambda _{1-n,-1} = Q \, \mu _{-n,-1} + \eta \, \mu _{-n,0}$ and the following gauge reducibility  
\begin{align*} 
\delta _{g+1} \mu _{-g,p} = Q \, \mu_{-1-g,p} + \eta \, \mu _{-1-g,p+1} \, , \hspace{5mm} ( g \geq n , \, p \in \mathbb{Z} ) \, . 
\end{align*}
If we also impose $\delta _{n+1} \mu _{-n,p} \in {\rm Ker}[Q \eta ]$ as $\mu _{-n,p} \in {\rm Ker}[Q \eta ]$, the next gauge parameters $\{ \mu _{-n-1,p} \} _{p}$ must live in ${\rm Ker}[Q \eta ]$. 
Then, a half part of the set of fields--antifields is given by 
\begin{align*}
\big{\{ } \Psi _{1-g,-1} \big{\} }_{g=0}^{n} \subset \cH _{\beta \gamma } \, , \hspace{5mm} 
\big{\{ } \Psi _{-g,p} \big{\} }_{g \geq n , p \in \mathbb{Z}} \subset {\rm Ker} [ Q \eta ] \, . 
\end{align*}
These fields must satisfy the constraint equations 
\begin{align*}
\eta \, \Psi _{1-g,-1} = 0 \, \hspace{3mm} (g <n ) \, ,\hspace{3mm} 
Q \, \Psi _{-n,-1} = \eta \, \Psi _{-n,0} \, , \hspace{3mm} 
\eta \, \Psi _{-g,p} + Q \, \Psi _{-g,p-1} = 0 \, \hspace{3mm} ({\rm otherwise}) \, . 
\end{align*}
If we do not restrict $\delta _{n+1} \mu _{-n,p}$, the higher gauge parameters $\{ \mu _{-g,p} \} _{g>n,p\in \mathbb{Z}}$ belong to the large Hilbert space $\cH _{\xi \eta \phi }$. 
Then, a half part of the set of fields--antifields is given by 
\begin{align*}
&\big{\{ }  \Psi _{1-g,-1} \big{\} }_{g=0}^{n} \subset \cH _{\beta \gamma } \, , \hspace{5mm} 
\big{\{ } (\Psi _{-n,p} )^{\ast } \big{\} }_{p \in \mathbb{Z}} \subset {\rm Ker}[Q \eta ] \, , \hspace{5mm} 
\big{\{ } (\Psi _{-g,p} )^{\ast } \big{\} }_{g>n , p \in \mathbb{Z}} \subset \cH  _{\xi \eta \phi } \, . 
\end{align*}
In this case, only the fields $\Psi _{g,p}$ labeled by $g \geq -n$ satisfy the constraint equations 
\begin{align*}
\eta \, \Psi _{1-g,-1} = 0 \, \hspace{3mm} (g <n ) \, ,\hspace{3mm} 
Q \, \Psi _{-n,-1} = \eta \, \Psi _{-n,0} \, , \hspace{3mm} 
\eta \, \Psi _{-n,p} + Q \, \Psi _{-n,p-1} = 0 \, \hspace{3mm} (p \in \mathbb{Z} ) \, , 
\end{align*}
and the other fields are constraint free. 
In either cases, there is no constraints on the other half of the set of fields--antifields, 
\begin{align*}
\big{\{ } (\Psi _{1-g,-1} )^{\ast } \big{\} }_{g=0}^{n} \subset \cH  _{\xi \eta \phi } \, , \hspace{5mm} 
\big{\{ } ( \Psi _{-g,p} )^{\ast } \big{\} }_{g \geq 0, p\in \mathbb{Z} } \subset \cH  _{\xi \eta \phi } \, . 
\end{align*}
The master action is given by the same form as the large form given in section 2 except for that the $g$- and $p$-labels run over the appropriate regions.

\end{document}